\documentclass{ieeeaccess}

\usepackage{cite}
\usepackage{amsmath,amssymb,amsfonts}
\usepackage{algorithmic}
\usepackage{graphicx}
\usepackage{textcomp}

\usepackage[hyphens]{url}

\usepackage{pifont}

\usepackage{siunitx}
\usepackage{microtype}

\usepackage{hyperref}

\definecolor{alizarin}{rgb}{0.82, 0.1, 0.26}

\newcommand{\orcid}[1]{\href{https://orcid.org/#1}{\includegraphics[width=8pt]{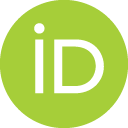}}}

\def\BibTeX{{\rm B\kern-.05em{\sc i\kern-.025em b}\kern-.08em
    T\kern-.1667em\lower.7ex\hbox{E}\kern-.125emX}}

\usepackage{draftwatermark}
\SetWatermarkText{This is the \\ prepublication \\ version}
\SetWatermarkScale{0.6}

\begin{document}

\history{Received September 12, 2021, accepted October 1, 2021, date of publication October 27, 2021.}
\doi{10.1109/ACCESS.2021.3123504}

\title{Toward a Consistent Taxonomy for Scenario-Based Development and Test Approaches for Automated Vehicles: A Proposal for a Structuring Framework, a Basic Vocabulary, and its Application}

\author{\uppercase{Markus~Steimle}\,\orcid{0000-0001-8913-6980}\authorrefmark{1},
\uppercase{Till~Menzel}\,\orcid{0000-0003-3467-6917}\authorrefmark{1}\uppercase{, and Markus~Maurer}\,\orcid{0000-0002-5357-9701}\authorrefmark{1}}
\address[1]{Institute of Control Engineering, Technische Universität Braunschweig, 
	38106 Braunschweig, Germany (e-mail: \{steimle, menzel, maurer\}@ifr.ing.tu-bs.de)}
\tfootnote{This research is funded by the German Federal Ministry for Economic Affairs and Energy within the project ``SET Level -- Simulation-based development and testing of automated driving,'' a successor project to the PEGASUS project and a project in the PEGASUS family (German abbreviation of \textbf{S}imulationsbasiertes \textbf{E}ntwickeln und \textbf{T}esten von automatisiertem Fahren). The results presented in this publication reflect the authors' opinions and not necessarily the opinion of all project participants.}

\markboth
{Steimle \headeretal: Toward a Consistent Taxonomy for Scenario-Based Development and Test Approaches for Automated Vehicles}
{Steimle \headeretal: Toward a Consistent Taxonomy for Scenario-Based Development and Test Approaches for Automated Vehicles}

\corresp{Corresponding author: Markus Steimle (e-mail: steimle@ifr.ing.tu-bs.de).}

\begin{abstract}
Ensuring and validating the safe operation of automated vehicles are key challenges for their market launch.
Scenario-based development and test approaches are currently being pursued as possible solutions.
An essential prerequisite for researching, applying, and standardizing these approaches is a consistent and agreed-upon taxonomy.
This taxonomy must include relevant terms, their descriptions, and the relationships between the respective terms. 
To the best of our knowledge, such a taxonomy does not yet exist, and this absence often leads to misunderstandings in, for example, coordination processes and discussions.
This publication contributes to this taxonomy. 
For this purpose, we propose a framework for structuring the taxonomy.
Within this framework, we propose a basic vocabulary by identifying and describing terms that we consider particularly relevant to an overview of such scenario-based development and test approaches.
Additionally, we visualize the proposed terms and the relationships between them as UML diagrams and explain the application of the proposed basic vocabulary using an example.
\end{abstract}


\titlepgskip=-15pt

\maketitle

\section{Introduction} \label{sec_introduction}

\PARstart{A}{utomated} vehicles have been regarded as a central research domain in the automotive industry for decades~\cite{Kroger_2016}.
The challenges involved in delivering automated vehicles to market include not only developing the vehicles but, in particular, validating their safe operation. 
Traditional distance-based safety validation, as applied to driver assistance systems, is inapplicable due to time and cost constraints. 
Wachenfeld and Winner~\cite{Wachenfeld_2016b} refer to it as an ``approval trap'' for autonomous driving.
However, scenario-based development and test approaches may be possible solutions to ensure and validate the safe operation of automated vehicles (e.g.,~\cite{Riedmaier_2020}, \cite{ISODIS21448_2021}, \cite{Schuldt_2017}, \cite{Wood_2019}, and~\cite{Stellet_2020}). 
These approaches have been investigated in recently completed research projects such as aFAS~\cite{Stolte_2015}, \mbox{ENABLE-S3}~\cite{ENABLE_S3_Method_2019}, and PEGASUS~\cite{PEGASUSMethod_2019}.
These scenario-based approaches are currently being further pursued and investigated in ongoing research projects such as SET Level~\cite{SL45_Homepage} and VVM~\cite{VVM_Homepage}, which are successors to the PEGASUS project and projects in the PEGASUS family.

To research, apply, and standardize these scenario-based development and test approaches, a consistent taxonomy is essential.
This taxonomy must include relevant terms, their descriptions, and the relationships between them.
Furthermore, researchers and developers in the domain of automated vehicles must agree on this taxonomy.
Unfortunately, to the best of our knowledge, such a consistent and agreed-upon taxonomy does not yet exist.
As a result, misunderstandings often arise in, for example, coordination processes and discussions.
These misunderstandings increase the time spent and costs incurred.

Various documents and standards define the terms related to development and test approaches (e.g.,~\cite{Wood_2019}, \cite{ISTQB_2020}, \cite{Spillner_2014}, \cite{ISO26262_2018}, \cite{SAE_J3016_2018}, \cite{IEEE829_2008}, and \cite{ISO29119_2013}).
However, most publications focus only on specific subdomains within development or test approaches for automated vehicles or on specific subdomains within traditional development or test approaches for software products.
Within each of these domains, there are different activities and people with different backgrounds.
As a result, a considerable number of terms have evolved from these different domains, activities, and people. 
These terms have been defined in various publications.
For these reasons, most of these terms are not explicitly related to each other. 
Therefore, the terms and their definitions do not always match.
Additionally, we have identified that visualizing the relationships between the terms from traditional software development and test approaches as UML diagrams reveals inconsistencies and inaccuracies in the current standards\footnote{
For example, in the case of the ISO~26262~standard~\cite{ISO26262_2018}, which addresses the functional safety of road vehicles, we identified an inaccuracy between the definitions of the terms \emph{(hardware) component} and \emph{hardware part} because the two definitions refer to each other.
A component is a ``non-system level \textit{element} [...] that is logically or technically separable and is comprised of more than one \textit{hardware part} [...] or one or more \textit{software units} [...].''
A hardware part is a ``portion of a hardware \textit{component} [...] at the first level of hierarchical decomposition.''
In our opinion, there is a cyclic dependency between the two definitions.
Additionally, consider the following example, where there is only one level of hierarchical decomposition: a microcontroller (component) consists of a CPU (hardware part) and software code (consisting of several software units).
This decomposition leads to a component consisting of a CPU (hardware part) and software code (consisting of several software units).
This decomposition would be excluded by the definition of the term \emph{component} mentioned above.
}.
Moreover, due to the large number of related publications, terms in different publications sometimes have the same name but different meanings.
There are also different meanings within individual publications\footnote{
For example, according to the IEEE~829~standard~\cite{IEEE829_2008}, which addresses software testing, the term \emph{scenario} can have three different meanings.
First, a scenario is ``a description of a series of events that may occur concurrently or sequentially [adopted from IEEE~1362~standard~\cite{IEEE1362_1998}].''
Second, a scenario is ``an account or synopsis of a projected course of events or actions [adopted from IEEE~1362~standard~\cite{IEEE1362_1998}].''
Third, a scenario is ``commonly used for groups of test cases; synonyms are script, set, or suite.''
}.
Furthermore, some terms that have already been defined in traditional software development and test approaches cannot be applied to scenario-based approaches for automated vehicles because they have a different meaning in the respective domain\footnote{
For example, consider the three meanings of the term \emph{scenario} as defined in the IEEE~829~standard~\cite{IEEE829_2008} (see footnote~2). 
Unfortunately, none of these meanings corresponds to the meaning of a scenario in scenario-based development and test approaches for automated vehicles.
In these approaches, according to Ulbrich~\textit{et~al.}~\cite{Ulbrich_2015b}, ``a scenario describes the temporal development between several scenes in a sequence of scenes. Every scenario starts with an initial scene. Actions \& events as well as goals \& values may be specified to characterize this temporal development in a scenario. Other than a scene, a scenario spans a certain amount of time.''}.

Therefore, this publication contributes to a consistent taxonomy for scenario-based development and test approaches for automated vehicles.
For this purpose, we propose a framework for structuring the taxonomy and a basic vocabulary within this framework.
We create the framework and basic vocabulary based on various requirements that we identify and based on a review of relevant literature.
The proposed basic vocabulary includes terms that we consider particularly relevant to an overview of such scenario-based development and test approaches.
We identify which terms are already defined in scenario-based development and test approaches for automated vehicles, which terms from traditional software development and test approaches can be transferred or adapted to these scenario-based approaches, and which terms are additionally necessary.
However, this publication focuses on scenario-based test approaches and does not aim to provide a complete taxonomy.
Since the domain of scenario-based development and test approaches for automated vehicles is interdisciplinary and currently very dynamic, this basic vocabulary will also serve as a basis for discussion.
Furthermore, the created framework should enable, for example, researchers and developers in the domain of automated vehicles to extend the proposed basic vocabulary and thus the taxonomy by adding further terms.
This contribution is based on two German publications that we presented at a conference on automated and connected driving~\cite{Steimle_2018},~\cite{Steimle_2018b}.
For discussion purposes, an earlier version of this publication was published on arXiv in April~2021~\cite{Steimle_2021}.
Due to this prepublication, the feedback we received is discussed and considered in this version.

\subsection*{Novelty and main contribution to the state of the art} 

The novelty and main contribution of this publication is a consistent taxonomy for scenario-based development and test approaches for automated vehicles.
Therefore, we

\begin{itemize}
 \item identify requirements that the taxonomy should fulfill.
 \item review the relevant literature.
 \item propose a framework for structuring the taxonomy.
 \item propose a basic vocabulary within the taxonomy by identifying and describing terms that we consider particularly relevant.
 \item identify the relationships between the terms that are part of the basic vocabulary.
 \item visualize the terms that are part of the basic vocabulary and their relationships using UML diagrams.
  \item explain the application of the basic vocabulary using an example.
 \item evaluate the current state of the taxonomy based on the identified requirements.
\end{itemize}

\subsection*{Structure}

This publication is structured as follows: 
Section~\ref{sec_requirements} presents the requirements that the taxonomy for scenario-based development and test approaches for automated vehicles should fulfill.
Section~\ref{sec_framework} describes the proposed framework for structuring this taxonomy.
Section~\ref{sec_terminology} presents the proposed basic vocabulary within the proposed framework.
To create this basic vocabulary, we propose and describe terms that we consider particularly relevant to an overview of these approaches. 
To provide a graphical presentation of the relationships between the proposed terms, we visualize the terms and their relationships using UML diagrams. 
We apply the proposed basic vocabulary in Section~\ref{sec_application} by explaining and distinguishing the proposed terms based on the development and testing of an adaptive cruise control (ACC) system. 
For this purpose, we provide examples for each proposed term.
Section~\ref{sec_req_evaluation} evaluates the current state of the taxonomy based on the identified requirements.
Finally, Section~\ref{sec_conclusion} discusses our conclusions and ideas for future work.

\section{Requirements for the taxonomy} \label{sec_requirements}

In this section, we present the requirements that the taxonomy for scenario-based development and test approaches for automated vehicles should fulfill.
We identified the following requirements:

\begin{itemize}
    \item The taxonomy should be as compliant as possible to existing standards~(Req.~1\,a), to the prevailing parlance in the scientific literature~(Req.~1\,b) and to the colloquial language of researchers and developers~(Req.~1\,c).

    \item The taxonomy should be clearly structured~(Req.~2\,a), clearly formulated (Req.~2\,b), defined as simple as possible~(Req.~2\,c) and defined without inconsistencies and contradictions~(Req.~2\,d).

   \item The taxonomy should be visualized clearly, concisely and understandably~(Req.~3).

    \item The taxonomy should be extensible and adaptable~(Req.~4) (see also \textit{note~1} below).

     \item The proposed basic vocabulary~--~including the description of the terms it contains and their relationships~--~representing the current state of the taxonomy should be as universally applicable as possible~(Req.~5) (see also \textit{note~2} below).

\end{itemize}

In the following paragraphs, we provide two notes on the requirements listed above.

\textit{Note~1 (belonging to Req.~4)}:
Thompson and Smith~\cite{Thompson_2007} and Hamburg~\cite{Hamburg_2014} addressed representing terms and their relationships as UML diagrams.
However, they represented only small fragments of terms and questioned the feasibility of a fixed UML diagram.
According to Hamburg~\cite{Hamburg_2014}, a generic UML diagram would be desirable.
In particular, concerning the International Software Testing Qualifications Board (ISTQB)~\cite{ISTQB_2020}, Hamburg~\cite{Hamburg_2014} is certain that different people can arrive at different results.
These inconsistencies occur because the ISTQB is not a single method but a collection of knowledge (see Section~\ref{sec_framework}). 
There may be different opinions regarding the selection of the \emph{correct} terms from the numerous testing terms. 
We agree with Hamburg's~\cite{Hamburg_2014} opinion.
We also believe that it is questionable whether there is a fixed taxonomy~--~visualized by a UML diagram~--~that contains all terms and their relationships that is also suitable for all use cases.
Furthermore, the domain of scenario-based development and test approaches for automated vehicles is currently very dynamic.
The considerations mentioned above will likely require future extensions and adaptations of the taxonomy.
Terms need to be adapted, new terms need to be added, and the relationships between terms need to be adapted.
Therefore, it should be possible for researchers and developers to extend and adapt the taxonomy according to their needs.

\textit{Note~2 (belonging to Req.~5)}:
This requirement is especially relevant to this publication.
As mentioned above, it is questionable whether there exists a fixed taxonomy that contains all terms and their relationships that is also suitable for all use cases.
Therefore, we focus on the terms and relationships that we consider particularly relevant to an overview of scenario-based development and test approaches for automated vehicles.
As mentioned earlier, these terms represent the proposed basic vocabulary.
The current state of the taxonomy proposed in this publication includes this basic vocabulary, the descriptions of the terms that are part of this basic vocabulary, and the relationships between these terms.
By focusing on particularly relevant terms and their relationships, the proposed taxonomy should be as universally applicable as possible.
This current state of the taxonomy serves as a basis for discussion and should be extended in the future by adding further terms. 
Therefore, a particular requirement within this publication is that the proposed basic vocabulary, including the description of the terms it contains and their relationships, should be as universally applicable as possible.

\section{Proposal of a framework for structuring the taxonomy for scenario-based development and test approaches for automated vehicles} \label{sec_framework}

To create a comprehensive taxonomy for scenario-based development and test approaches for automated vehicles, it is essential to structure the taxonomy.
In this section, we propose and describe a framework for structuring this taxonomy.
In our opinion, a generic development and test process is an appropriate starting point for creating such a framework.
Orientation toward a development and test process allows us to build the taxonomy systematically.
Additionally, we can divide the taxonomy into different phases and, therefore, into smaller parts that can be analyzed and presented separately. 
This subdivision helps to focus on terms that we consider particularly relevant to an overview of such scenario-based development and test approaches within each phase.
Furthermore, this subdivision leads to an overview of a basic vocabulary but also allows for the possibility of adding further terms and therefore extending the taxonomy. 
Additionally, this subdivision simplifies the understanding and extension of the taxonomy, as someone can focus on specific phases.
By structuring the taxonomy according to a generic development and test process, the taxonomy is easy to understand because these processes are widely known.
However, this publication intends to describe the development and test process not in detail but only to the extent necessary to provide a framework to structure the taxonomy and derive the basic vocabulary.

\subsection{Overview of the Proposed Framework} \label{ssec_framework_overview}

Fig.~\ref{fig_framework} shows an overview of the proposed framework for structuring the taxonomy for scenario-based development and test approaches for automated vehicles.
The creation of this framework is described below.

In the automotive industry, for example, the ISO~26262 standard~\cite{ISO26262_2018} is used.
This standard is a guideline for developing safety-critical electrical/electronic systems and thus also defines a framework for developing automated vehicles according to aspects of functional safety.
In our opinion, the V-model-based process outlined in this standard is an appropriate starting point for creating a framework that structures the taxonomy for scenario-based development and test approaches for automated vehicles.
As the name ``scenario-based'' indicates, scenarios have a special meaning in both scenario-based development and test approaches for automated vehicles.
According to Menzel~\textit{et~al.}~\cite{Menzel_2018a}, scenarios can be used in several steps of the V-model process outlined in the ISO~26262 standard~\cite{ISO26262_2018} to generate the required work products.
For this purpose, ``scenarios can be identified on a high level of abstraction in the concept phase and be detailed and concretized along the development process.''
The derived scenarios can be used during the test process as, for example, a basis for test cases to validate the safe behavior of automated vehicles.

In summary, we propose dividing the taxonomy into three primary parts, as illustrated in Fig.~\ref{fig_framework}: concept, design, and implementation (left branch of the V-model); the test process (right branch of the V-model); and scenario derivation and usage.
Although the latter is illustrated as a parallel part, it is not a parallel phase but is linked to the other visualized phases' activities.
Since our focus is on scenario-based test approaches, we propose a finer subdivision of the test process in the next subsection.

\Figure[t!](topskip=0pt, botskip=0pt, midskip=0pt)[]{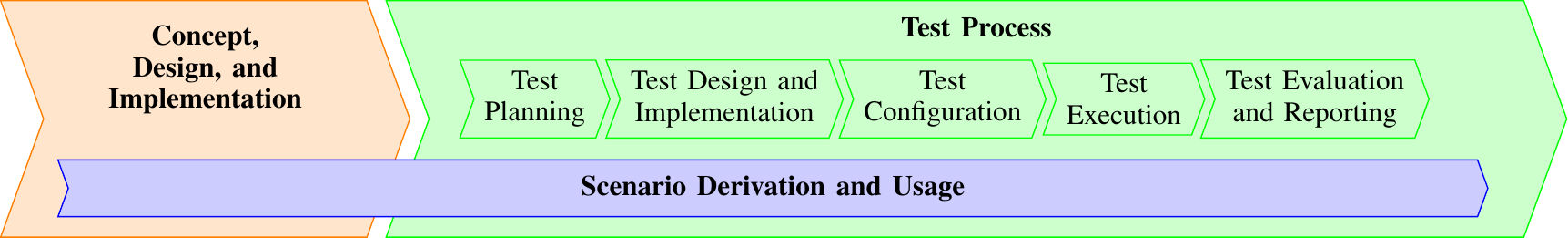}
{Overview of the proposed framework for structuring the taxonomy for scenario-based development and test approaches for automated vehicles. Although the scenario derivation and usage is illustrated as a parallel part, it is not a parallel phase but is linked to the other visualized phases' activities.\label{fig_framework}}

\subsection{Detailing the Test Process as Part of the Proposed Framework} \label{ssec_framework_test_process}

As this publication focuses on scenario-based test approaches, we propose a finer subdivision of the test process based on the test phases contained therein.
As a basis for this detailing, relevant aspects for testing automated vehicles are briefly described below.

Saust~\textit{et~al.}~\cite{Saust_2009} focused on tests of autonomous vehicles in urban environments during their development.
According to Saust~\textit{et~al.}~\cite{Saust_2009}, a modular design of the overall system is a prerequisite to achieving the necessary test depth.
Only then can individual modules be independently developed, commissioned, and tested.
This modularization, which is standard in the automotive industry, motivates testing at different levels. 
Examples of these levels include the vehicle level, the system level, and the component level.

The development process outlined in the ISO~26262 standard~\cite{ISO26262_2018} provides the prerequisite for modularity, and individual modules can be developed and tested at different levels.
Depending on the test level considered, tests can be scenario-based or based on simple tests without scenarios.
According to the ISO~26262 standard~\cite{ISO26262_2018}, in the automotive domain, product development is performed at four levels that correspond to the test levels: the vehicle level, the system level, the component level, and the unit level. 
In the traditional software domain, the vehicle level does not exist because there is no vehicle.
In this domain, a distinction is made between the system level, the component level (also called the integration level), and the unit level (also called the module level). 
According to Hoffmann~\cite{Hoffmann_2013}, 
system tests assume that the software has been tested nearly exclusively from a functional perspective (black-box testing).
In contrast, integration and module tests focus on the internal code structure (white-box testing). 
However, the clear dividing line drawn in the literature is violated in practice in many places~\cite{Hoffmann_2013}.

Independent of the exact division of these test levels, an independent (sub)test process consisting of test phases with defined test objects and test objectives can be defined at each test level.
According to Witte~\cite{Witte_2016} and Spillner~\textit{et~al.}~\cite{Spillner_2014}, the individual test phases should not be performed strictly sequentially even though they are often illustrated sequentially, as overlaps in time cannot be avoided in practice. 
The individual test phases should provide only a rough framework for orientation. 
In our opinion, this rough framework can be used to structure the taxonomy related to the test process of scenario-based development and test approaches for automated vehicles.
This structure allows us to divide the entire test process into smaller units to analyze and visualize them separately. 
In the literature, different numbers and different names of test phases are used. 
Examples of publications with the number of different test phases considered are listed below:

\begin{itemize}
	
	\item ISO/IEC/IEEE~29119~\cite{ISO29119_2013}: 2 main test phases, 7 detailed test phases in total
	
	\item IEEE~829~(1998)~\cite{IEEE829_1998}: 3 main test phases, 8 detailed test phases in total
	
	\item IEEE~829~(2008)~\cite{IEEE829_2008}: 4 test phases
	
	\item ISTQB~\cite{ISTQB_2020}: 5 test phases (until 03/2018); 7 test phases (since 04/2018)
	
	\item Spillner~\textit{et~al.}~\cite{Spillner_2014}: 5 test phases	

	\item Horstmann~\cite{Horstmann_2005}: 7 test phases	
	
	\item Witte~\cite{Witte_2016}: 6 test phases
\end{itemize}

The test phases outlined in the publications referenced above have different names due to the different numbers and different aspects of the test phases considered. 
In summary, however, they cover similar or identical aspects. 
As mentioned above, this publication does not aim to describe a detailed development and test process.
For this reason, we do not provide a detailed description of the test phases considered in the publications referenced above.
For a detailed description of the considered test phases, interested readers are referred to the corresponding publications.

Based on our research and experience gained during the design of the taxonomy as well as to provide structure to the taxonomy, we propose the test phases described below for inclusion within the test process of the proposed framework.
The proposed phases are only intended to structure the taxonomy and not to explain an (extensive) test process.
Test processes may be divided into different phases by different companies, and they may have different names and consist of different activities.
Therefore, a formal definition of these test phases is not within the scope of this publication.
A short summary of the activities and artifacts that are part of the proposed test phases is provided below. 
A detailed description of these activities and artifacts is provided in the respective subsections regarding the test phases in Section~\ref{sec_terminology}.

In all analyzed publications, test planning is performed at the beginning of the test process; therefore, we propose \textbf{test planning} as the first test phase.
This phase includes the creation of test plan(s).
The next test phase is the \textbf{test design and implementation} phase, where the test specification is created.
The test specification specifies, among other things, the test cases.
Subsequently, the elements necessary to execute the specified test cases are defined.
We call this test phase the \textbf{test configuration} phase. 
In the subsequent \textbf{test execution} phase, the test cases are executed.
The test case evaluation can be performed simultaneously with the test execution but also as an independent activity based on recorded data after the test execution.
Therefore, we propose considering it a separate test phase. 
We call this test phase the \textbf{test evaluation and reporting} phase.
As mentioned above, these test phases should not be performed strictly sequentially even if they are illustrated sequentially; rather, they  provide a rough framework for orientation.
That is, the activities belonging to the test phases can overlap in time and can also be carried out iteratively, with some activities and documents being revisited.
The proposed test phases are illustrated in Fig.~\ref{fig_framework}.
Again, we note that Fig.~\ref{fig_framework} aims not to present a comprehensive development and test process but to provide a framework for structuring the taxonomy.

To structure the taxonomy and derive the basic vocabulary related to the test process, we concentrate on the ISTQB’s activities. 
In a presentation, Thompson and Smith~\cite{Thompson_2007} showed that the terms related to testing have gradually been developed in the extensive literature.
The ISTQB has contributed to an agreement of these terms, so focusing on the ISTQB activities seems reasonable.
Furthermore, according to a statement on its homepage, the ISTQB has created the most successful scheme for the certification of software testers worldwide~\cite{ISTQBHomepage}.
For this reason, many software testers should be familiar with the test phases and terms used there.
Additionally, we concentrate on the ISO/IEC/IEEE~29119~standard~\cite{ISO29119_2013} since it bundles and replaces different standards such as the IEEE~829~standard~\cite{IEEE829_2008}, the IEEE~1008~standard~\cite{IEEE1008_1987}, and the BS~7925~standard~\cite{BS7925_1998} (Part~1: Vocabulary and Part 2: Software component testing).
For completeness, we would like to mention that the ISO/IEC/IEEE~29119~standard~\cite{ISO29119_2013} is somewhat controversial.
For example, some people think that the standard is too extensive and that the community was not involved in its development.
However, this discussion does not influence the proposed structure of the taxonomy or the derivation of the basic vocabulary.

\section{Proposal of a basic vocabulary for scenario-based development and test approaches for automated vehicles} \label{sec_terminology}

Based on the proposed framework illustrated in Fig.~\ref{fig_framework}, in the following three subsections, we analyze the different phases and propose terms and their descriptions that we consider particularly relevant to an overview of scenario-based development and test approaches for automated vehicles.
We evaluate which terms are already defined in scenario-based development and test approaches for automated vehicles, which terms can be transferred or adapted from traditional software testing to such approaches, and which additional terms are required.
In this way, we propose a basic vocabulary for a consistent taxonomy. 
We present only the terms and dependencies that we consider particularly relevant.
Most of these terms have dependencies on additional terms that are not part of the proposed basic vocabulary and therefore are not illustrated and described in detail. 
These additional terms will be added to the proposed basic vocabulary to extend the taxonomy in the future.

The following subsections are structured as follows: 
At the beginning of each subsection, we provide a short description of the activities, contents, and special aspects of the respective phase. 
Then, we refer to a UML diagram that visualizes the terms we propose for the basic vocabulary and their relationships. 
Finally, we describe the terms illustrated in the associated UML diagram (if they have not been described in one of the previous subsections). 
The terms are printed in bold and marked with arrows ({\color{alizarin}\Pifont{pzd}{\char227}}) at the position where they are described.
For a complete overview of the taxonomy, the individual UML diagrams can be combined into one large UML diagram.
Due to space limitations, this combined presentation is not included in this publication.

To represent the relationships and dependencies between the proposed terms (and the artifacts they represent) in the form of UML diagrams, we use associations, aggregations, and compositions by which the respective terms are connected.
The type of relationship (described textually) and the associated multiplicity are additionally provided for each individual connection. 
An association (represented by a line) generally relates two terms.
An arrow indicates the navigation direction. 
To show existence dependencies between terms, we use aggregations and compositions.
The creation process of the respective artifacts represented by the terms is also considered to represent whether the child artifact exists when the associated parent artifact does not (yet) exist.
A composition (represented by a line and a filled diamond) indicates that the child artifact only exists if the parent artifact (marked with a diamond) exists. 
For example, if the parent artifact represents a building that consists of several rooms (children), the rooms only exist if the building exists. 
In comparison, an aggregation (represented by a line and an unfilled diamond) indicates that the child artifact can exist independently of the parent artifact.
For more details on the UML notation, we refer interested readers to the \emph{OMG Unified Modeling Language Publication}~\cite{UML_2017}.

\subsection{Scenario Derivation and Usage} \label{ssec_scenarioDerivation}

As the name \emph{scenario-based development and test approaches} indicates, the central term in scenario derivation and usage is the term \emph{scenario} and related terms.
Based on Ulbrich~\textit{et~al.}~\cite{Ulbrich_2015b}, the ISO~21448 draft international standard~\cite{ISODIS21448_2021} addresses the term \emph{scenario} and its related terms and proposes corresponding definitions.
As mentioned above, according to Menzel~\textit{et~al.}~\cite{Menzel_2019}, ``scenarios can be identified on a high level of abstraction in the concept phase and be detailed and concretized along the development process.''
For this purpose, Menzel~\textit{et~al.}~\cite{Menzel_2018a} extended the definition of a scenario proposed by Ulbrich~\textit{et~al.}~\cite{Ulbrich_2015b}, using three abstraction levels to represent scenarios: functional, logical, and concrete.
Functional scenarios are described in language and are human-readable. 
Logical scenarios detail functional scenarios on a state-space level, and concrete scenarios concretize logical scenarios.
The derived scenarios can be used during the test process as a basis for test cases to, for example, validate whether an automated vehicle's behavior is safe.
For an exact allocation of functional, logical, and concrete scenarios to the different steps in the development and test process, see Menzel~\textit{et~al.}~\cite{Menzel_2018a}.
Neurohr~\textit{et~al.}~\cite{Neurohr_2021} extended these three abstraction levels through abstract scenarios, which are situated between the functional and logical abstraction levels.
According to Neurohr~\textit{et~al.}~\cite{Neurohr_2021}, ``an abstract scenario is a formalized, declarative description of a traffic scenario focusing on complex relations, particularly on causal relations.''
Since this work is still in progress, we currently neglect abstract scenarios in the basic vocabulary and propose its future inclusion.

According to a discussion with Fuchs~\textit{et~al.}~\cite{Fuchs_2021} regarding scenarios and test cases and based on our own experience, there is often a discussion regarding whether evaluation criteria are part of the scenario.
In our opinion, a scenario should not contain evaluation criteria. 
These evaluation criteria should be part of the test case, which, following our suggestion, consists of a scenario and one or more evaluation criteria. 
A more detailed explanation of the term \emph{test case} is provided in Section~\ref{ssec_TestSpecification}.

Since most of the terms and definitions related to a scenario are widely accepted according to our analysis and experience, we propose adopting these definitions for the basic vocabulary, except for the modifications proposed below.
One modification we propose is to rename the term \emph{dynamic elements}, used by Ulbrich~\textit{et~al.}~\cite{Ulbrich_2015b} and the ISO~21448 draft international standard~\cite{ISODIS21448_2021}, to \emph{movable objects}.
This modification is meant to emphasize that an object can be temporarily static at a certain point in time within the scenario even though it will move at a later point in time; examples of movable objects include a car that temporarily stops at a traffic light or a parked car that will move at a later point in time. 
Both objects are not dynamic at a considered point in time but will move later during the execution of the scenario.
As an additional example, we consider an aluminum can that may be part of the stationary 3-D world if it is not supposed to move during the execution of the scenario.  
In this case, the can is part of the scenery.
However, if the can is supposed to move during the scenario, it must be specified as a movable object in the scenario specification.
In contrast, if the can is part of the stationary 3-D world (i.e., the scenery) in a simulation, it can never move during the execution of that scenario.
Thus, depending on the specific scenario, an object can be part of the scenery or a movable object.
These details must be considered when specifying the scenario.
An additional term that we think is particularly relevant to include in the basic vocabulary is \emph{real-world test drive}, which has not previously been defined.

Fig.~\ref{fig_scenario_UML} visualizes the terms related to a scenario that we consider particularly relevant to an overview of scenario-based development and test approaches for automated vehicles and their relationships as a UML diagram. 
In the following paragraphs, we describe the terms illustrated in this UML diagram.

\Figure[t!](topskip=0pt, botskip=0pt, midskip=0pt)[]{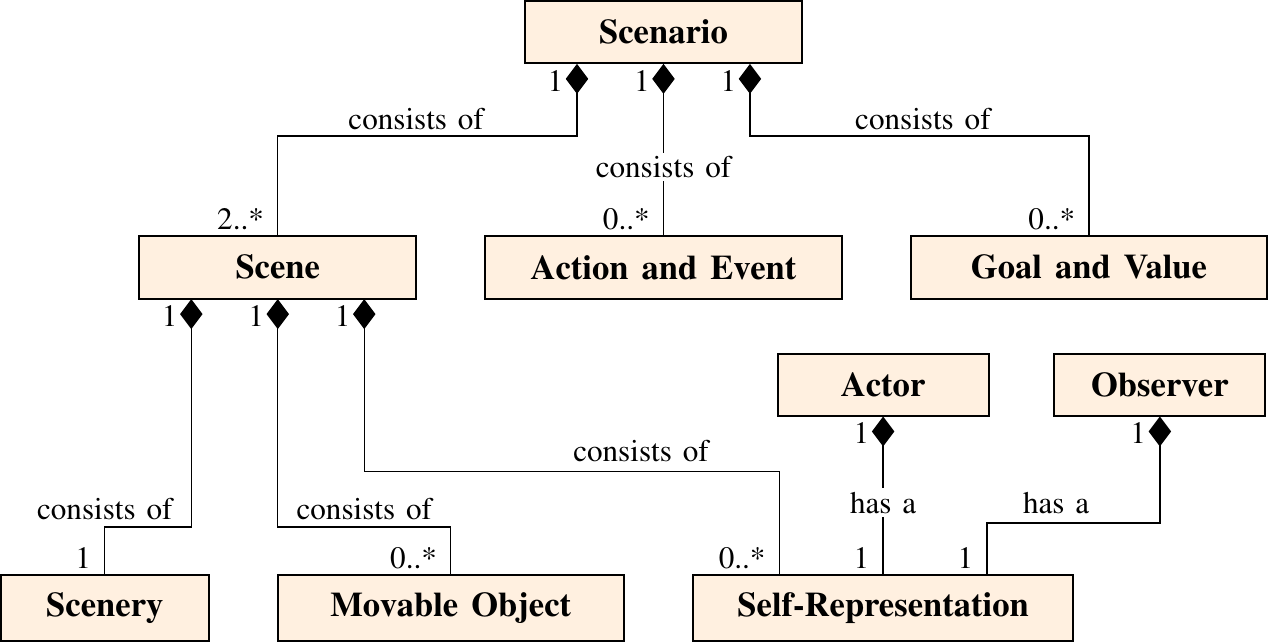}
{Terms proposed for the basic vocabulary related to a scenario visualized as a UML diagram.\label{fig_scenario_UML}}

	{\color{alizarin}\Pifont{pzd}{\char227}} 
	\textbf{Scenario}: 
	``A scenario describes the temporal development between several scenes in a sequence of scenes.
	Every scenario starts with an initial scene. 
	Actions \& events as well as goals \& values may be specified to characterize this temporal development in a scenario. 
	Other than a scene, a scenario spans a certain amount of time.''~\cite{Ulbrich_2015b}

	{\color{alizarin}\Pifont{pzd}{\char227}} 
	\textbf{Scene}: 
	``A scene describes a snapshot of the environment including the scenery and [movable objects]\footnote{``dynamic elements'' was used in the original text; an explanation of the replacement of this term with ``movable objects'' is found in the text above.}, as well as all actors’ and observers’ self-representations, and the relationships among those entities. 
	Only a scene representation in a simulated world can be all-encompassing (objective scene, ground truth).
	In the real world[,] it is incomplete, incorrect, uncertain, and from one or several observers’ points of view (subjective scene).''~\cite{Ulbrich_2015b}

	{\color{alizarin}\Pifont{pzd}{\char227}}
	\textbf{Scenery}: 
	``The scenery subsumes all geo-spatially stationary aspects of the scene. 
	This entails metric, semantic, and topological information about roads and all their components like lanes, lane markings, road surfaces, or the roads’ domain types.
	Moreover, this subsumes information about conflict areas between lanes as well as information about their interconnections, for example, at intersections. 
	Apart from the before mentioned environment conditions, the scenery also includes stationary elements like houses, fences, curbs, trees, traffic lights, or traffic signs.''~\cite{Ulbrich_2015b}

	{\color{alizarin}\Pifont{pzd}{\char227}}
	\textbf{Movable object}: A movable object is an object that moves according to kinetic energy or that is supposed to move within the scenario. 
	The respective object can be temporarily static.
	
	{\color{alizarin}\Pifont{pzd}{\char227}}
	\textbf{Actor}: 
	``An actor is an element of a scene acting on its own behalf.''~\cite{Ulbrich_2015b}

	{\color{alizarin}\Pifont{pzd}{\char227}}
	\textbf{Observer}: 
	``An observer\textsuperscript{1} is a perceiving element within the scene or is observing the scene as a whole. [...]
	\textsuperscript{1}:\,This is not an observer as in the sense of control engineering.''~\cite{Ulbrich_2015b}

	{\color{alizarin}\Pifont{pzd}{\char227}}
	\textbf{Self-representation}: The self-representation of an actor or observer contains its current skills and abilities as well as its states and attributes~\cite{Ulbrich_2015b}. 
	According to Nolte~\textit{et~al.}~\cite{Nolte_2020}, the self-representation of an actor or observer\footnote{In the original text, Nolte~\textit{et~al.}~\cite{Nolte_2020} used the more general term ``system''. Since this description refers to an actor or observer, the general term ``system'' has been replaced with ``actor or observer''.} ``\textit{consists of the set of explicit internal models of the} [\textit{actor's or observer's}] \textit{properties. 
	This set of models allows the} [\textit{actor or observer}] \textit{to infer knowledge about its own logic and dynamic state and to assess its own possible actions.}''	
	An overview of the models that are part of the actor's or observer's self-representation is provided in~\cite{Nolte_2020}.

	{\color{alizarin}\Pifont{pzd}{\char227}}
	\textbf{Action and event}:
	An action is an atomic behavior performed by an actor in a scene.	
	An event is an occurrence at a particular point in time~\cite{ISODIS21448_2021}.

	{\color{alizarin}\Pifont{pzd}{\char227}}
	\textbf{Goal and value}:
	Actors have individual goals and values. 
	In contrast to actions and events, a scenario can also be described solely by an initial scene and the command to all actors to pursue their individual goals and values without specifying any further scenes.
	Through these goals and values, actions and events are independently determined and executed by the assigned actors. 
	Goals and values can be transient (e.g.,~mission or operator commands) or permanent (e.g., regulatory or societal).~\cite{Ulbrich_2015b}

Fig.~\ref{fig_scenario_levels_UML} visualizes the terms related to the different abstraction levels of scenarios that we consider particularly relevant to an overview of scenario-based development and test approaches for automated vehicles and their relationships as a UML diagram. 
In the following paragraphs, we describe the terms illustrated in this UML diagram, which have not been described thus far.

\Figure[t!](topskip=0pt, botskip=0pt, midskip=0pt)[]{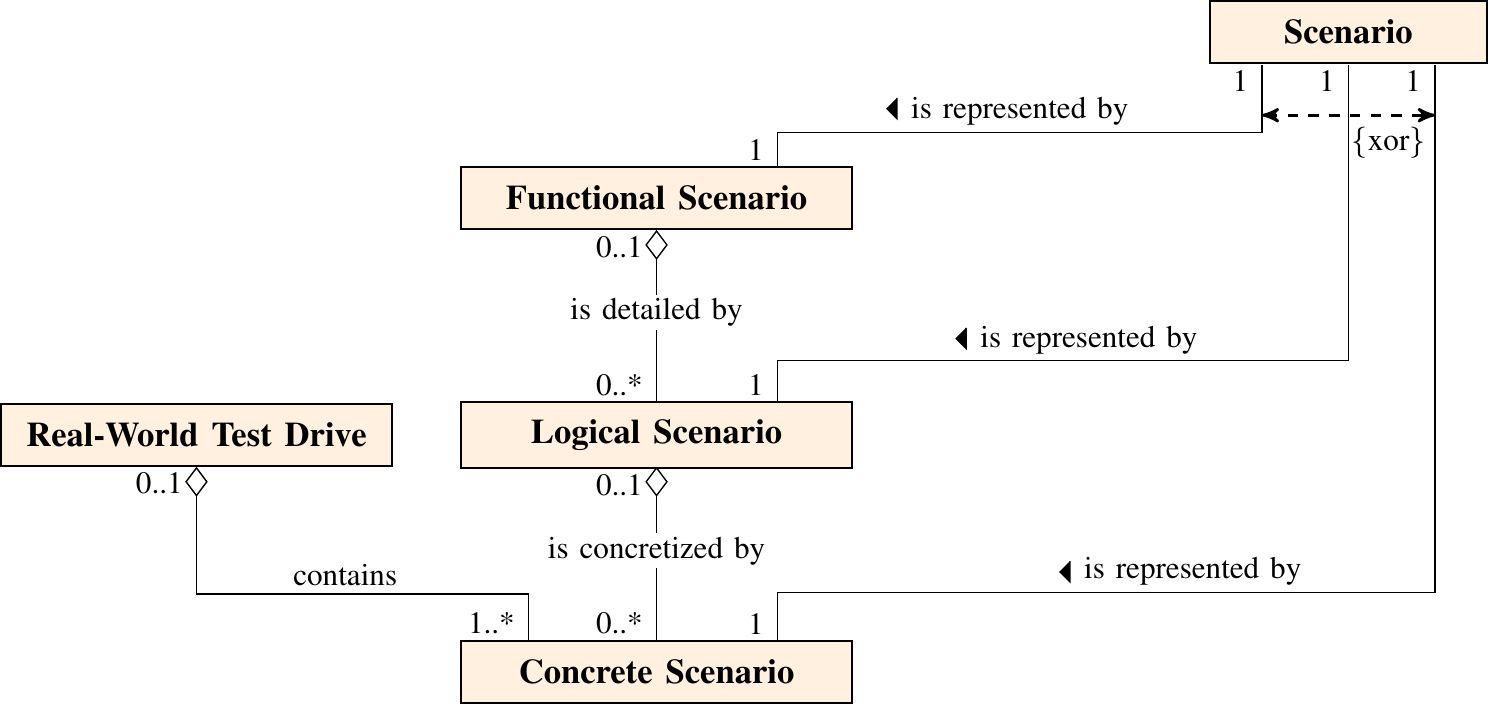}
{Terms proposed for the basic vocabulary related to different abstraction levels of scenarios visualized as a UML diagram.\label{fig_scenario_levels_UML}}

	{\color{alizarin}\Pifont{pzd}{\char227}} 
 	\textbf{Functional scenario}: 
	``Functional scenarios include operating scenarios on a semantic level.
	The entities of the domain and the relations of those entities are described via a linguistic scenario notation. 
	The scenarios are consistent. 
	The vocabulary used for the description of functional scenarios is specific for the use case and the domain and can feature different levels of detail.''~\cite{Menzel_2018a}

	{\color{alizarin}\Pifont{pzd}{\char227}} 
	\textbf{Logical scenario}: 
	``Logical scenarios include operating scenarios on a state-space level. 
	Logical scenarios represent the entities and the relations of those entities with the help of parameter ranges in the state space. 
	The parameter ranges can optionally be specified with probability distributions. 
	Additionally, the relations of the parameter ranges can optionally be specified with the help of correlations or numeric conditions.
	A logical scenario includes a formal notation of the scenario.''~\cite{Menzel_2018a}

	{\color{alizarin}\Pifont{pzd}{\char227}} 
	\textbf{Concrete scenario}: 
	``Concrete scenarios distinctly depict operating scenarios on a state-space level. 
	Concrete scenarios represent entities and the relations of those entities with the help of concrete values for each parameter in the state space.''~\cite{Menzel_2018a}

	{\color{alizarin}\Pifont{pzd}{\char227}} 
	\textbf{Real-world test drive}:
	A real-world test drive is a drive using a vehicle in the real world carried out for testing purposes.
	A real-world test drive contains one or more concrete scenarios, each of which can be assigned to one or more predefined logical scenarios.

\subsection{Concept, Design, and Implementation Phase} \label{ssec_DesignImplementation}

In the concept, design, and implementation phase, the product is developed (left branch of the V-model).
Since our focus is on scenario-based test approaches, only the terms most relevant to the basic vocabulary of such test approaches are described below.
For a detailed description of the concept, design, and implementation phase, please refer to, for example, the ISO~26262 standard~\cite{ISO26262_2018}.

The development of an automated vehicle includes, among other things, the development of one or more vehicle functions. 
A vehicle function represents a vehicle's functional behavior that will be implemented by one or more items and that is observable by the customer. 
An item is specified by an item definition.
According to the ISO~26262 standard~\cite{ISO26262_2018}, the objectives of an item definition are ``to define and describe the item, its functionality, dependencies on, and interaction with, the driver, the environment and other items at the vehicle level.''
An item's functionality will be implemented by the respective item and can be partially described using functional scenarios.
These functional scenarios can be detailed throughout the development process by logical scenarios and concretized by concrete scenarios. 
These different scenario types are described in Section~\ref{ssec_scenarioDerivation}. 
Additionally, the item definition can be used as a basis for deriving requirements.
According to the ISO~26262 standard~\cite{ISO26262_2018}, an item consists of one or more systems consisting of one or more components.
A component consists of possible hardware parts and/or possible software units.
A hardware part, in turn, consists of hardware subparts, which consist of hardware elementary subparts.
Regarding the definitions provided in the ISO~26262 standard~\cite{ISO26262_2018}, we propose the following modifications.

Regarding the term \emph{component}, we propose to supplement the definition listed in the ISO~26262 standard~\cite{ISO26262_2018} to the effect that a component can consist of possible subcomponents.
Furthermore, this definition of the term \emph{component} includes the subsentence ``[...] is comprised of more than one \textit{hardware part} [...] or one or more \textit{software units} [...].''
Regarding this definition, consider the following example in which there is only one level of hierarchical decomposition: a microcontroller (component) consists of a CPU (hardware part) and software code (consisting of several software units).
Thus, the component consists of one hardware part and several software units. 
This decomposition would be excluded by the definition of the term \emph{component} listed in the ISO~26262~standard~\cite{ISO26262_2018}.
Therefore, we propose to generalize the subsentence mentioned above by changing it to ``[...] is comprised of [one or more]\footnote{``more than one'' was used in the original text; an explanation of the replacement of this phrase with ``one or more'' is provided in the text above.} \textit{hardware part}[\textit{s}] [and/]or one or more \textit{software units}.''

Regarding the definition of the terms \emph{(hardware) component} and \emph{hardware part} provided in the ISO~26262 standard~\cite{ISO26262_2018}, we identified that both definitions refer to each other, which is why a cyclic dependency exists. 
To resolve this cyclic dependency, we propose adapting the definition of the term \emph{hardware part} as well as the definitions of related terms.
According to our proposal, a hardware part is a hardware element that is logically or technically separable at the first level of hierarchical decomposition.
A hardware part consists of possible hardware subparts if more than one level of hierarchical decomposition is considered during the analysis.
Therefore, a hardware subpart is a hardware element that is logically or technically separable at the second or greater level of hierarchical decomposition.
A hardware element that is logically or technically separable at the lowest level of hierarchical decomposition considered in the analysis is called a hardware elementary subpart.

Based on a discussion with Fuchs~\textit{et~al.}~\cite{Fuchs_2021} regarding the different levels mentioned above, we would like to point out that classifying a specific element into one of these levels depends on the developer and cannot be determined in general. 
For example, for a vehicle manufacturer, an item can be a complete vehicle and a system can be an ACC system, whereas for a supplier, the ACC system can also be an item. 
Thus, it cannot be generally said that an ACC system must always be at the system level. 
However, this ambiguity does not contradict the proposed basic vocabulary but must be considered when decomposing the item and applying the taxonomy.

Regarding the definition of the term \emph{item}, we propose to generalize the definition provided in the ISO~26262 standard~\cite{ISO26262_2018} by removing the phrase ``to which ISO~26262 is applied,'' as also suggested in the ISO~4804~technical report~\cite{ISO4804_2020}.

\Figure[t!](topskip=0pt, botskip=0pt, midskip=0pt)[]{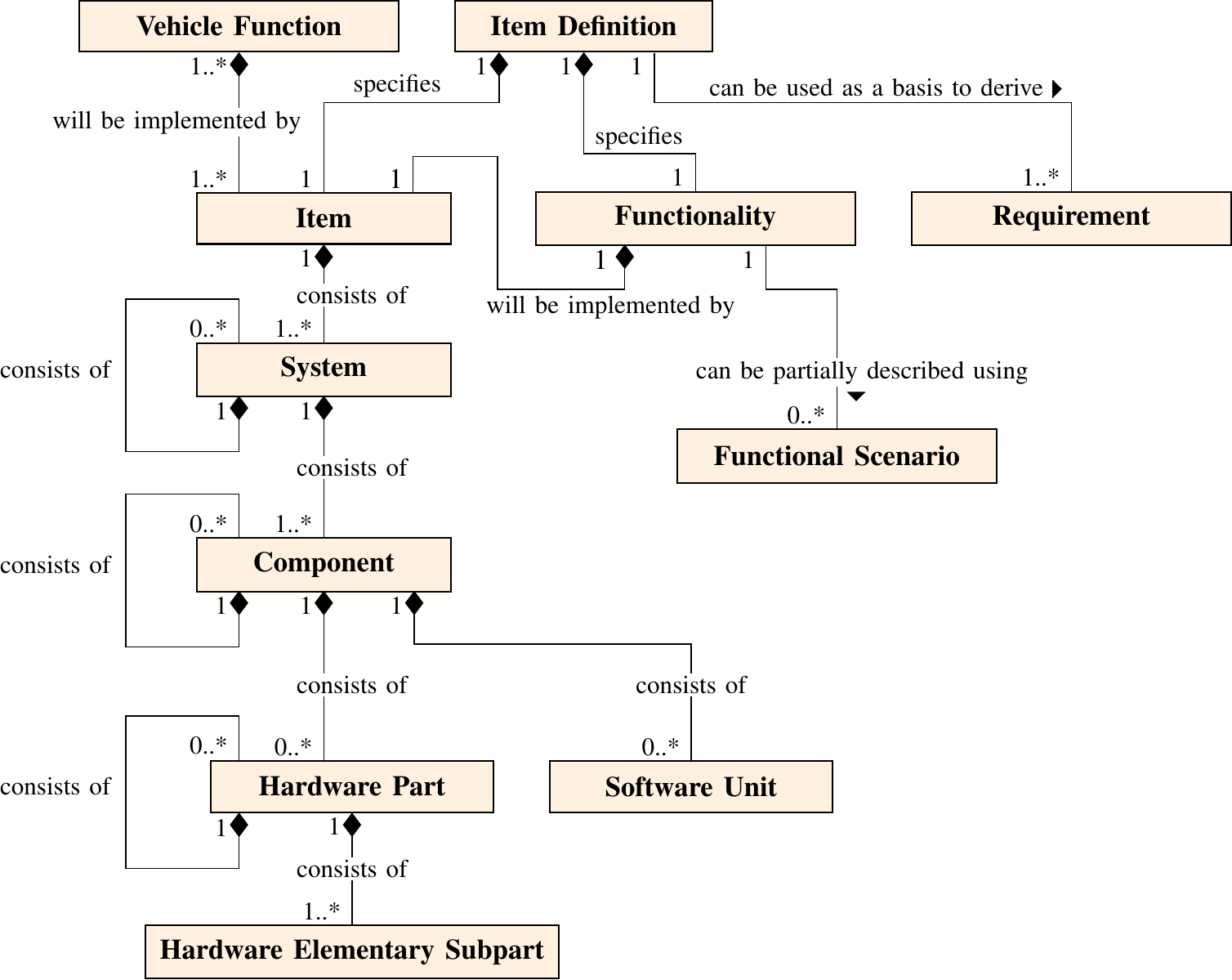}
{Terms proposed for the basic vocabulary related to the concept, design, and implementation phase visualized as a UML diagram.\label{fig_ProductDevelopment_UML}}

Fig.~\ref{fig_ProductDevelopment_UML} visualizes the terms related to the concept, design, and implementation phase that we consider particularly relevant to an overview of scenario-based development and test approaches for automated vehicles and their relationships as a UML diagram. 
Regarding this UML diagram, the ISO~26262 standard~\cite{ISO26262_2018} proposes a similar diagram in part~10 (Guidelines on ISO~26262) in Section~4.2, about which we want to list a few comments and propose some improvements.
In the UML diagram depicted in the ISO~26262 standard~\cite{ISO26262_2018}, only the terms \emph{item}, \emph{function}, \emph{system}, \emph{component}, and \emph{HW-part/SW-unit} are listed. 
We propose to include at least the directly related terms listed in the ISO~26262 standard~\cite{ISO26262_2018}, such as the term \emph{hardware elementary subpart}.
Furthermore, the ISO~26262 standard~\cite{ISO26262_2018} uses only aggregations and realizations. 
Concerning this type of representation, we propose using compositions to represent existence dependencies as well. 
Additionally, in the UML diagram we created, we textually describe the type of relationship and the associated multiplicity for each connection at both ends.
These additions increase understanding of the relationships between these terms; the ISO~26262 standard~\cite{ISO26262_2018} does not describe the type of relationship at this level of detail and only specifies the multiplicity at one end of the connection.
In addition, at the lowest level of the UML diagram depicted in the ISO~26262 standard~\cite{ISO26262_2018}, the multiplicity for \emph{HW-Part/SW-Unit} is denoted as 1..*; in our opinion, this multiplicity should be 0..* since at a certain level of hierarchical decomposition, a component can only consist of several subcomponents and may not necessarily contain hardware parts or software units at this level. 
At a lower level, these subcomponents may then consist of hardware parts and software units. 
This decomposition would be excluded in the UML diagram shown in the ISO~26262 standard~\cite{ISO26262_2018}.
In summary, the representation of the terms as a UML diagram in the ISO~26262 standard~\cite{ISO26262_2018} is excellent.
However, in our opinion, it would be helpful if this diagram was not shown in part~10 and was instead shown directly in part~1 of the ISO~26262 standard~\cite{ISO26262_2018}, which addresses vocabulary.
In the following paragraphs, we describe the terms illustrated in Fig.~\ref{fig_ProductDevelopment_UML}, which have not been described thus far.

 	{\color{alizarin}\Pifont{pzd}{\char227}} 
	\textbf{Vehicle function}:
	A vehicle function represents the vehicle's functional behavior, which will be implemented by one or more items and is observable by the customer~\cite{ISO26262_2018}.

	{\color{alizarin}\Pifont{pzd}{\char227}} 
	\textbf{Item definition}:
	The item definition is a document that specifies ``the item, its functionality, dependencies on, and interaction with, the driver, the environment and other items at the vehicle level.''~\cite{ISO26262_2018}

	{\color{alizarin}\Pifont{pzd}{\char227}} 
	\textbf{Item}: 
	An item is a ``\textit{system} [...] or combination of \textit{systems} [...]\footnote{``to which ISO 26262 is applied'' was present in the original text; an explanation of its deletion is included in the text above.} that implements a function or part of a function at the vehicle level.''~\cite{ISO26262_2018}

	{\color{alizarin}\Pifont{pzd}{\char227}} 
	\textbf{Functionality}:
	A functionality (specified in an item definition) represents the vehicle's functional behavior, which will be implemented by the corresponding item.
	The functionality of an item can be partially described using functional scenarios.

	{\color{alizarin}\Pifont{pzd}{\char227}} 
	\textbf{Requirement}:
	Requirements can be derived based on the item definition.
	According to the ISO/IEC/IEEE~29148 standard~\cite{ISO29148_2018}, a requirement is ``a statement which translates or expresses a need and its associated \textit{constraints} [...] and \textit{conditions} [...]. [...] Requirements exist at different levels in the system structure. [...] A requirement is an expression of one or more particular needs in a very specific, precise and unambiguous manner. [...] A requirement always relates to [an item,] system, software or service, or other item of interest.''
	A requirement is verifiable by an evaluation criterion.

	{\color{alizarin}\Pifont{pzd}{\char227}} 
	\textbf{System}: 
	A system is a ``set of \textit{components} [...] or subsystems that relates at least a sensor, a controller and an actuator with one another. [...]
	The related sensor or actuator can be included in the system, or can be external to the system.''~\cite{ISO26262_2018}

	{\color{alizarin}\Pifont{pzd}{\char227}} 
	\textbf{Component}:
	A component is a ``non-system level \textit{element} [...] that is logically or technically separable and is comprised of [one or more]\footnote{``more than one'' was used in the original text; an explanation of the replacement of this phrase with ``one or more'' is provided in the text above.} \textit{hardware part}[\textit{s}] [...] [and/]or one or more \textit{software units} [...].''~\cite{ISO26262_2018}
	A component can consist of possible subcomponents.
	If a component consists of only one or more hardware parts, it can also be referred to as a hardware component. 
	If a component consists of only one or more software units, it can also be referred to as a software component.~\cite{ISO26262_2018}

	{\color{alizarin}\Pifont{pzd}{\char227}} 
	\textbf{Software unit}: 
	A software unit is an ``atomic level \textit{software component} [...] of the software \textit{architecture} [...] that can be subjected to stand-alone \textit{testing} [...].''~\cite{ISO26262_2018}

	{\color{alizarin}\Pifont{pzd}{\char227}} 
	\textbf{Hardware part}: 
	A hardware part is a hardware element that is logically or technically separable at the first level of hierarchical decomposition.
	A hardware part consists of possible hardware subparts if more than one level of hierarchical decomposition is considered in the analysis.

	{\color{alizarin}\Pifont{pzd}{\char227}} 
	\textbf{Hardware elementary subpart}: 
	A hardware elementary subpart is a hardware element that is logically or technically separable at the lowest level of hierarchical decomposition considered in the analysis.

\subsection{Test Process} \label{ssec_TestProcess}

In this subsection, we propose and describe terms related to the test process that we consider particularly relevant to an overview of scenario-based development and test approaches for automated vehicles.
As mentioned above, we structured the taxonomy related to the test process according to the test phases shown in Fig.~\ref{fig_framework}.
The proposed terms in each test phase are visualized as a UML diagram in the corresponding test phase.

\subsubsection{Test Planning Phase} \label{ssec_TestPlanning}

The test process starts with the test planning phase.
In this test phase, test plan(s) are created.
According to our research, the test planning phase in scenario-based development and test approaches for automated vehicles is essentially similar to that in traditional software testing, which is described in detail in, for example, the ISO/IEC/IEEE~29119~standard~\cite{ISO29119_2013}.
Since the ISO/IEC/IEEE~29119~standard~\cite{ISO29119_2013} is the most current and comprehensive standard, we propose adopting the basic vocabulary terms using this standard, except for the modifications proposed below.

The ISO/IEC/IEEE~29119~standard~\cite{ISO29119_2013} uses the term \emph{test item} in most definitions.
However, according to its glossary entry, the term \emph{test item} is interchangeable with the term \emph{test object}. 
Since the term \emph{item} is already used in the ISO~26262~standard~\cite{ISO26262_2018}, we propose using the term \emph{test object} instead of the term \emph{test item} to avoid confusion.
According to a discussion with Fuchs~\textit{et~al.}~\cite{Fuchs_2021} regarding the term \emph{test object}, terms such as \emph{software under test}, \emph{device under test}, or \emph{vehicle under test} are often used instead of the term \emph{test object}.
Since these terms already classify the type of test object (e.g., a piece of software, a device, or a vehicle), we propose to use the superordinate term \emph{test object} if no specific type of test object is intended.
A detailed description of the term \emph{test object} is provided in the list at the end of this subsection.

The ISO/IEC/IEEE~29119~standard~\cite{ISO29119_2013} is partially limited to software products in terms of the definitions listed therein. 
However, since scenario-based development and test approaches for automated vehicles are not limited to software products, we propose to generalize the definitions provided in the ISO/IEC/IEEE~29119~standard~\cite{ISO29119_2013} by removing the word ``software'' in the corresponding definitions.

The terms \emph{sub-process test plan}, \emph{test sub-process plan}, and \emph{test sub-process test plan} are used inconsistently and interchangeably in the ISO/IEC/IEEE~29119~standard~\cite{ISO29119_2013} but each has the same meaning. 
To maintain consistency, we propose the uniform use of the term \emph{sub-process test plan}.
A detailed description of the term \emph{sub-process test plan} is provided in the list at the end of this subsection.

Regarding the term \emph{test objective}, the literature contains various definitions, which range from very general to very specific.
In a very general definition, the test objective is the reason or purpose of testing, according to the ISTQB~\cite{ISTQB_2020}.
The ISO~29119 standard~\cite{ISO29119_2013} only states that a test is performed to achieve one or more test objectives; it does not provide a specific definition for the term \emph{test objective}.
Spillner~\textit{et~al.}~\cite{Spillner_2014} divided the test objectives into different test levels and adapted them to the focus of the respective test levels. 
Therefore, the test objectives are specific to each test level.
In component testing, the test objective is to check ``that the entire functionality of the test object works correctly and completely as required by its specification. Here, functionality means the input/output behavior of the test object.''~\cite{Spillner_2014}
In the integration test, the test objective is to detect ``interface problems as well as conflicts between integrated parts.''~\cite{Spillner_2014}
In the system test, the test objective is to verify ``whether the complete system meets the specified functional and non-functional requirements [...] and how well it does that. Failures from incorrect, incomplete, or inconsistent implementation of requirements should be detected. Even undocumented or forgotten requirements should be identified.''~\cite{Spillner_2014}
According to the descriptions mentioned above, it is important that test objectives can be divided into different test levels. 
For this purpose, test objectives can be structured hierarchically and specified at different levels of detail.
A detailed description of the term \emph{test objective} is provided in the list at the end of this subsection.

Fig.~\ref{fig_TestPlanning_UML} visualizes the terms related to the test planning phase that we consider particularly relevant to an overview of scenario-based development and test approaches for automated vehicles and their relationships as a UML diagram. 
In this subsection, we explicitly point out that the terms shown in Fig.~\ref{fig_TestPlanning_UML} represent only a small part of all terms related to the test planning phase.
The terms listed in Fig.~\ref{fig_TestPlanning_UML} are those we consider particularly relevant to an overview, and they will be used in the following subsections. 
For a detailed explanation of all activities and artifacts related to the test planning phase, we refer interested readers to, for example, the ISO/IEC/IEEE~29119~standard~\cite{ISO29119_2013}. 
In the following paragraphs, we describe the terms illustrated in the corresponding UML diagram, which have not been described thus far.

\Figure[t!](topskip=0pt, botskip=0pt, midskip=0pt)[]{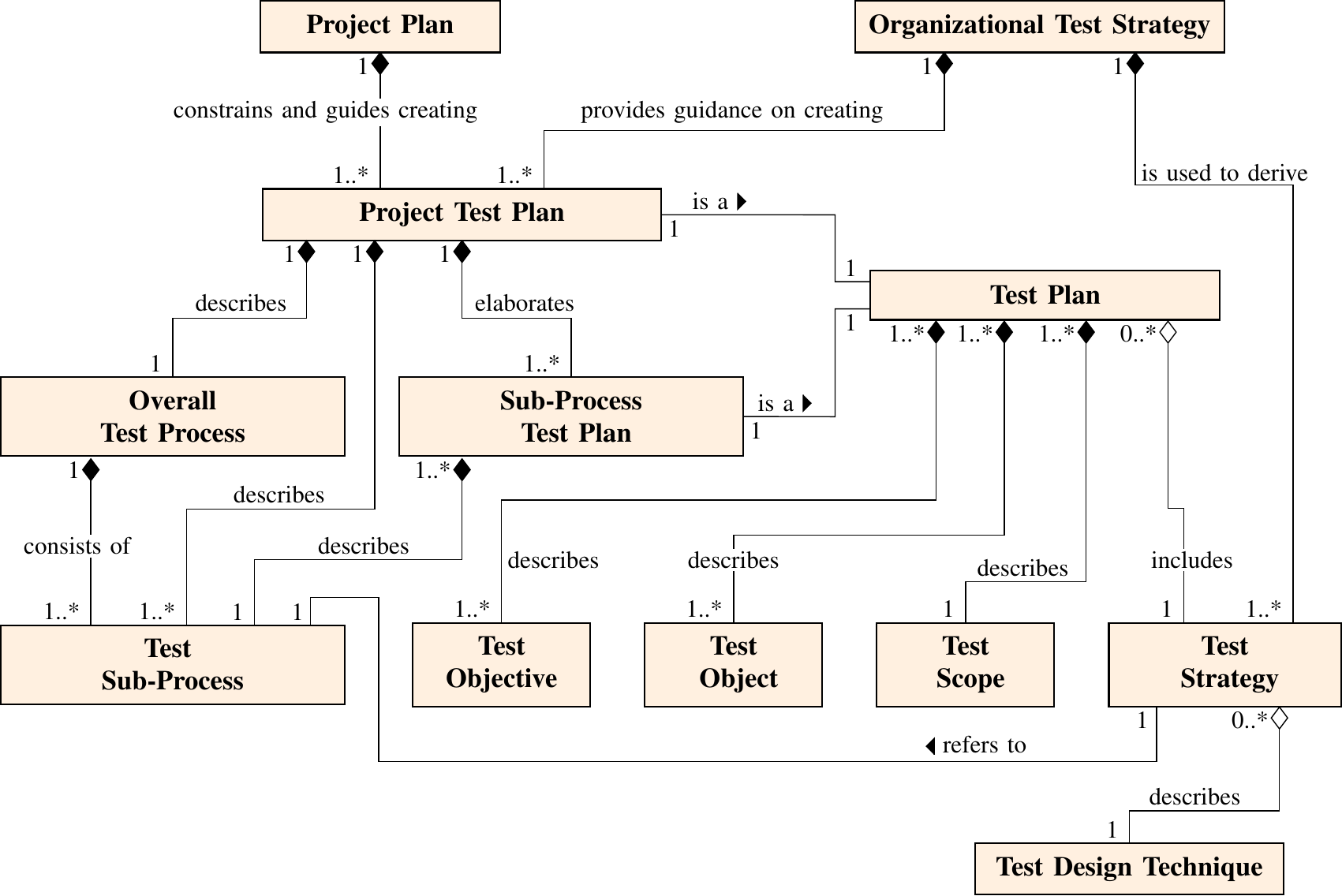}
{Terms proposed for the basic vocabulary related to the test planning phase visualized as a UML diagram.\label{fig_TestPlanning_UML}}

{\color{alizarin}\Pifont{pzd}{\char227}} 
\textbf{Project plan}:
The project plan is a ``formal, approved document used to guide both project execution and project control. 
The primary uses of the project plan are to document planning assumptions and decisions, facilitate communication among \textit{\textbf{stakeholders}}, and document approved scope, cost, and schedule \textit{\textbf{baselines}}. 
A project plan may be summary or detailed.'' 
Thus, the project plan constrains and guides the creation of the project test plan.~\cite{ProjectManagementInstitute_2000}

{\color{alizarin}\Pifont{pzd}{\char227}} 
\textbf{Organizational test strategy}:
The organizational test strategy is a ``document that expresses the generic requirements for the testing to be performed on all the projects run within the organization, providing detail on how the testing is to be performed.''
Thus, the organizational test strategy provides guidance on creating the project test plan.~\cite{ISO29119_2013}

{\color{alizarin}\Pifont{pzd}{\char227}} 
\textbf{Project test plan}:
A project test plan ``describes the overall strategy for testing and the test processes to be used. 
It establishes the context of testing for the project by determining the objectives, practices, resources, and schedule; it also identifies the applicable test sub-processes (e.g. system testing, performance testing).''~\cite{ISO29119_2013}

{\color{alizarin}\Pifont{pzd}{\char227}} 
\textbf{Overall test process}:
The (overall) test process ``provides information on the quality of a [...]\footnote{``software'' was present in the original text; an explanation of its deletion is provided in the text above.} product, often comprised of a number of activities, grouped into one or more test sub-processes. [...]
The Test Process for a particular project may well consist of multiple sub-processes, e.g. a system test sub-process, a Test Planning sub-process (a part of the larger Test Management Process) or a static testing sub-process.''~\cite{ISO29119_2013}

{\color{alizarin}\Pifont{pzd}{\char227}} 
\textbf{Test sub-process}:
A test sub-process consists of ``test management and dynamic (and static) test processes used to perform a specific test level (e.g. [component testing, integration testing,] system testing, acceptance testing) or test type (e.g. usability testing, performance testing) normally within the context of an overall test process for a test project. [...]
A test sub-process could comprise one or more test types. 
Depending on the life cycle model used, test sub-processes are also typically called test phases, test levels, test stages or test tasks.''~\cite{ISO29119_2013}

{\color{alizarin}\Pifont{pzd}{\char227}} 
\textbf{Sub-process test plan}:
A sub-process test plan is elaborated based on a project test plan.
``Each sub-process test plan may address more than one test level (e.g. the security test plan may address several test levels) and may address more than one test type (e.g. a System Test Plan that addresses functional and performance testing at the system test level).
The sub-process test plan will also describe the strategy for test execution (e.g. scripted, unscripted, or a mixture of both).''~\cite{ISO29119_2013}

{\color{alizarin}\Pifont{pzd}{\char227}} 
\textbf{Test plan}:
A test plan is a ``detailed description of test objectives to be achieved and the means and schedule for achieving them, organized to coordinate testing activities for some test [object] or set of test [objects]\footnote{``item'' was used in the original text; an explanation of the replacement of this term with ``object'' is provided in the text above.}. [...]
A project can have more than one Test Plan, for example there could be a Project Test Plan (also known as a master test plan) that encompasses all testing activities on the project; further detail of particular test activities could be defined in one or more [sub-process test plans]\footnote{``test sub-process plans'' was used in the original text; an explanation of the replacement of this term with ``sub-process test plans'' is provided in the text above.} (i.e. a system test plan or a performance test plan). [...]
Typically a Test Plan is a written document, though other plan formats could be possible as defined locally within an organization or project.''~\cite{ISO29119_2013}

{\color{alizarin}\Pifont{pzd}{\char227}} 
\textbf{Test objective}:
A test objective is a reason or purpose for deriving and executing test cases. 
Test objectives justify the chosen test scope.
They can be hierarchical and subdivided into further (sub)test objectives. 
Test objectives can be divided into different test levels and specified at different levels of detail.

{\color{alizarin}\Pifont{pzd}{\char227}} 
\textbf{Test object}:
The test object is the hardware and/or software to be tested and is specific to the considered test level.
Depending on the nature of the test object, it may also be referred to, for example, as a ``system under test,'' ``device under test,'' or ``vehicle under test''.

{\color{alizarin}\Pifont{pzd}{\char227}} 
\textbf{Test scope}:
The test scope ``summarizes the features of the test [object(s)] to be tested. [It] also identifies any features of the test [object(s)] that are to be specifically excluded from testing and the rationale for their exclusion\footnote{``item(s)'' was used in the original text; an explanation of the replacement of this term with ``object(s)'' is provided in the text above.}.''~\cite{ISO29119_2013}

{\color{alizarin}\Pifont{pzd}{\char227}} 
\textbf{Test strategy}:
The test strategy ``[...] describes the approach to testing for a specific test project or test sub-process or sub-processes. [...]
The test strategy usually describes some or all of the following: the test practices used; the test sub-processes to be implemented; the retesting and regression testing to be employed; the test design techniques and corresponding test completion criteria to be used; test data; test environment and testing tool requirements; and expectations for test deliverables.''~\cite{ISO29119_2013}

{\color{alizarin}\Pifont{pzd}{\char227}} 
\textbf{Test design technique}:
A test design technique consists of ``activities, concepts, processes, and patterns used to construct a test model that is used to identify test conditions for a test [object]\footnote{``item'' was used in the original text; an explanation of the replacement of this term with ``object'' is provided in the text above.}, derive corresponding test coverage items, and subsequently derive or select test cases.''~\cite{ISO29119_2013}

\subsubsection{Test Design and Implementation Phase} \label{ssec_TestSpecification}

The test design and implementation phase follows the test planning phase.
In this test phase, a test specification is created.
The starting points for creating the test specification (for scenario-based test approaches) are the specified scenarios, the corresponding test plan(s), and the requirements.

There are various definitions of the term \emph{test specification} and different definitions for the closely related terms \emph{test design specification}, \emph{test case specification}, and \emph{test procedure specification} in the literature.
Some authors, for example, Horstmann~\cite{Horstmann_2005} and Witte~\cite{Witte_2016}, use only the term \emph{test specification} and neglect the other three terms.
In our opinion, the distinction between the four terms is essential as used by, for example, the ISO/IEC/IEEE~29119~standard~\cite{ISO29119_2013}, Spillner~\textit{et~al.}~\cite{Spillner_2014}, the ISTQB~\cite{ISTQB_2020}, and the IEEE~829~standard~\cite{IEEE829_2008}.
Therefore, we propose adding the four specification-specific terms mentioned above to the basic vocabulary and adopting their definitions from the ISO/IEC/IEEE~29119~standard~\cite{ISO29119_2013}, which have also been adopted by the ISTQB~\cite{ISTQB_2020}.
Descriptions of these terms follow in the list at the end of this subsection.

In contrast to traditional software testing, we propose adapting the term \emph{test case} for scenario-based test approaches for automated vehicles. 
In these approaches, a test case consists of at least one concrete scenario (which does not exist in this form in traditional software testing) and one or more evaluation criteria.
In traditional software testing, e.g.,~according to the definition from the ISO/IEC/IEEE~29119~standard~\cite{ISO29119_2013}, which was adopted by the ISTQB~\cite{ISTQB_2020} in relation to content, a test case is a ``set of test case preconditions, inputs (including actions, where applicable), and expected results, developed to drive the execution of a test item to meet test objectives, including correct implementation, error identification, checking quality, and other valued information.''
This definition does not exactly correspond to the meaning of a test case in scenario-based test approaches for automated vehicles.
To emphasize this difference, we propose using the term \emph{concrete test case} instead of the more generic term \emph{test case} if there is a possibility of misunderstanding.
Furthermore, the term \emph{test scenario} is sometimes used instead of the proposed term, \emph{(concrete) test case}, in scenario-based test approaches for automated vehicles. 
We recommend avoiding the term \emph{test scenario}, as it has a different meaning in traditional software testing and can lead to misunderstandings. 
For example, the ISTQB~\cite{ISTQB_2020} and Spillner~\textit{et~al.}~\cite{Spillner_2014} define the term \emph{test scenario} as a set of test sequences, where a test sequence is defined as ``a set of several test cases in which the postcondition of one test is often used as the precondition for the next one.''
This definition of a test scenario does not reflect the meaning of a test case in scenario-based test approaches for automated vehicles.
Additionally, the term \emph{test scenario} does not clearly indicate whether it involves a concrete scenario (without evaluation criteria) from which test cases can be derived or a concrete scenario that is already combined with evaluation criteria.
Both uses can be found in the literature. 
In the absence of a prior description of the term \emph{test scenario}, this ambiguity can easily lead to misunderstanding or confusion.
For these reasons, we recommend avoiding the term \emph{test scenario} and using the term \emph{(concrete) test case} to indicate that it is a concrete scenario combined with evaluation criteria.

In contrast to traditional software testing, in the automotive industry, there are usually many different types of test benches (e.g., a software-in-the-loop, a hardware-in-the-loop, and a vehicle-in-the-loop test bench) with different characteristics that can be used to execute test cases.
Since different test benches may require, for example, different formats for specifying the test case specification and test procedure specification, it is necessary to determine the format in which they have to be specified before creating them.
For example, a test case specification created for a software-in-the-loop test bench must be machine-readable, and the OpenSCENARIO~\cite{OpenSCENARIO} and OpenDRIVE~\cite{OpenDRIVE} formats can be used to describe concrete scenarios, which are part of test cases.
To our knowledge, there is not yet an agreed-upon format for specifying test cases.
In contrast, a test case specification created for a human-operated test vehicle must also be human-readable.
Therefore, the test design specification must also specify the format of the test case specification and test procedure specification, which depends on the type of test bench planned to execute the test cases.

Compared to traditional software testing, scenario-based test approaches require the possibility that evaluation criteria are applied only within certain periods under certain circumstances~\cite{King_2019}.
For example, consider the verification of the automated deceleration of a vehicle by an ACC system, where the ACC system is activated at a specific point in time within the scenario.
In this scenario, the evaluation criterion addressing automated deceleration must only be applied when the ACC system is active.
Therefore, it must be possible to specify a corresponding application period for the evaluation criteria.

Fig.~\ref{fig_TestSpecification_UML} visualizes the terms related to the test design and implementation phase that we consider particularly relevant to an overview of scenario-based development and test approaches for automated vehicles and their relationships as a UML diagram. 
In the following paragraphs, we describe the terms illustrated in this UML diagram, which have not been described thus far.

\Figure[t!](topskip=0pt, botskip=0pt, midskip=0pt)[]{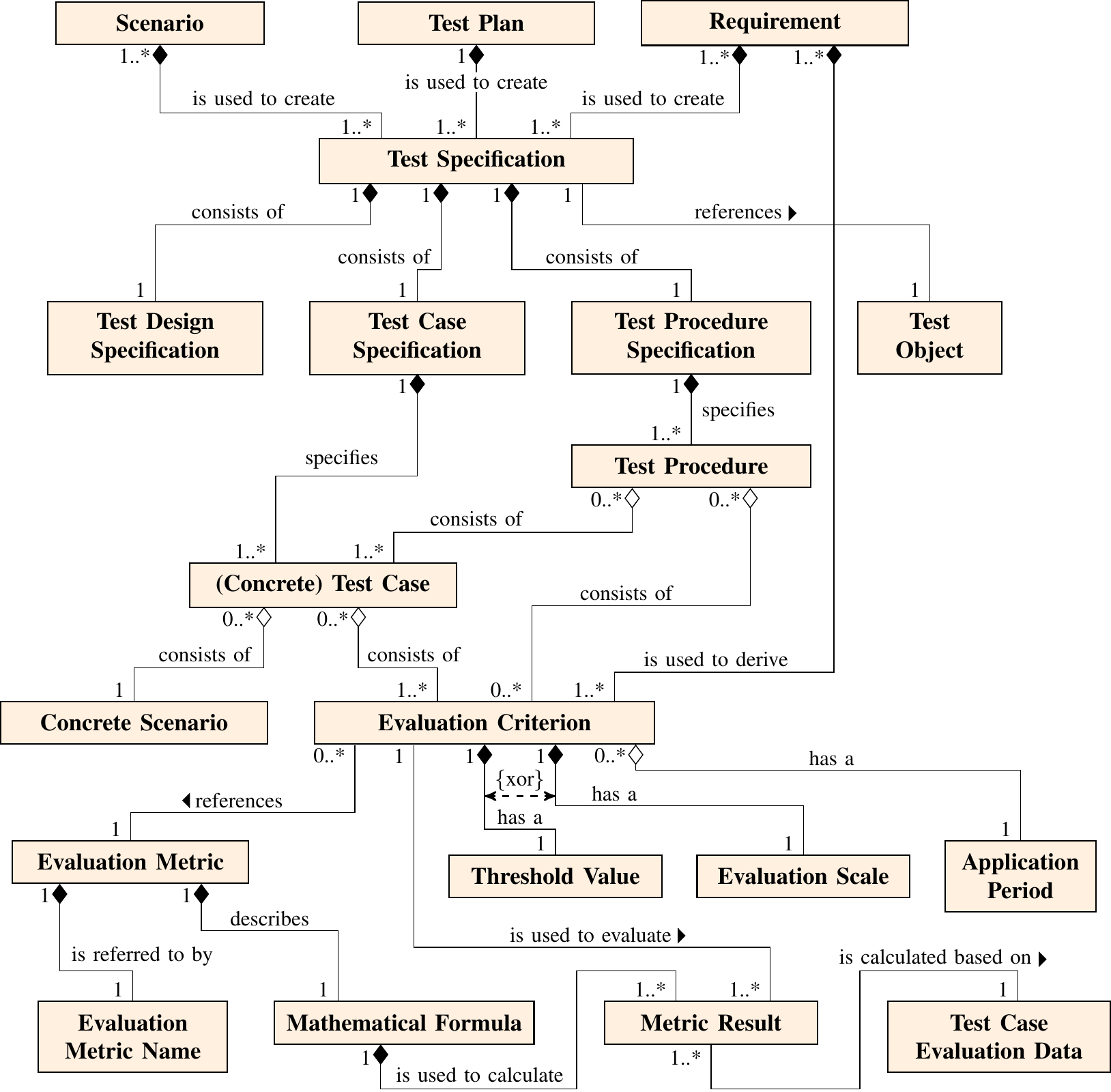}
{Terms proposed for the basic vocabulary related to the test design and implementation phase visualized as a UML diagram.\label{fig_TestSpecification_UML}}

{\color{alizarin}\Pifont{pzd}{\char227}} 
\textbf{Test specification}: The test specification is the ``complete documentation of the test design, test cases and test procedures for a specific test [object]\footnote{``item'' was used in the original text; an explanation of the replacement of this term with ``object'' is provided in the text above.}. [...]
A test specification could be detailed in one document, in a set of documents, or in other ways, for example in a mixture of documents and database entries.''~\cite{ISO29119_2013} 
Therefore, the test specification consists of the test design specification, the test case specification, and the test procedure specification for a specific test object.

{\color{alizarin}\Pifont{pzd}{\char227}} 
\textbf{Test design specification}: The test design specification is a document that specifies ``the features to be tested and their corresponding test conditions.''~\cite{ISO29119_2013} 
Additionally, the test design specification specifies the format of the test case specification and test procedure specification, which depend on the type of test bench planned to execute the test cases.

{\color{alizarin}\Pifont{pzd}{\char227}} 
\textbf{Test case specification}:
The test case specification is a ``documentation of a set of one or more test cases.''~\cite{ISO29119_2013}
The format used to specify the test case specification is described in the test design specification.

{\color{alizarin}\Pifont{pzd}{\char227}} 
\textbf{Test procedure specification}:
The test procedure specification is a document that specifies ``one or more test procedures, which are collections of test cases to be executed for a particular objective.''~\cite{ISO29119_2013}
The format used to specify the test procedure shall be described in the test design specification.
Additionally, the test procedure specification may also specify zero or more test bench configurations (see next subsection) that should be used to execute the test procedure(s).

{\color{alizarin}\Pifont{pzd}{\char227}} 
\textbf{Test procedure}:
A test procedure consists of ``a sequence of test cases in execution order, and [includes] any associated actions that may be required to set up the initial preconditions and any wrap up activities post execution.''~\cite{ISO29119_2013}
In addition to the evaluation criteria for individual test cases, further evaluation criteria can be defined for one or more test procedures. 
These additional evaluation criteria enable not only the individual evaluation of the executed test cases but also the cross-test-case or cross-test-procedure evaluation of one or more test procedures.
This cross-cutting evaluation can be used, for example, as a basis to statistically argue that an automated vehicle's behavior was safe within several concrete scenarios.

{\color{alizarin}\Pifont{pzd}{\char227}} 
\textbf{(Concrete) test case}:
A (concrete) test case consists of a concrete scenario and one or more evaluation criteria.
A (concrete) test case can be executed using different types of test benches, for example, with a software-in-the-loop test bench, a hardware-in-the-loop test bench, or a test vehicle driving on a proving ground.
If there is a possibility of misunderstanding, we propose using the term \emph{concrete test case} instead of the more general term \emph{test case}.

{\color{alizarin}\Pifont{pzd}{\char227}} 
\textbf{Evaluation criterion}:
An evaluation criterion is used to evaluate one or more metric results in relation to a threshold value or an evaluation scale within a specified application period.
These metric results are calculated using a mathematical formula (described by an evaluation metric) and data generated during the execution of the test case (test case evaluation data).
Thus, an evaluation criterion references an evaluation metric; it has a threshold value or an evaluation scale, and it has an application period. 
Evaluation criteria exist at different levels, which means that an evaluation criterion can be used to evaluate a single scene within a scenario, an entire scenario, a set of scenarios (i.e., a test procedure), or even a set of test procedures.

{\color{alizarin}\Pifont{pzd}{\char227}} 
\textbf{Evaluation metric}:
An evaluation metric is referred to by an evaluation metric name and describes a mathematical formula.
This formula is used to calculate one or more metric results based on the data generated during the execution of a test case (test case evaluation data).
Examples of evaluation metrics related to automated driving are the metric known as time-to-collision (TTC) and the metric known as post-encroachment-time (PET) (each includes an associated mathematical formula).

{\color{alizarin}\Pifont{pzd}{\char227}} 
\textbf{Evaluation metric name}:
An evaluation metric name (e.g., TTC or PET) refers to a specific evaluation metric used to calculate one or more associated metric results.

{\color{alizarin}\Pifont{pzd}{\char227}} 
\textbf{Mathematical formula}:
A mathematical formula (described by an evaluation metric) is a calculation rule used to convert input values (generated during the execution of a test case) at a specific point in time into an output value (called a metric result) that can be used to evaluate the test case.

{\color{alizarin}\Pifont{pzd}{\char227}} 
\textbf{Metric result}:
A metric result is calculated using a mathematical formula (described by an evaluation metric) and data (e.g., from single scenes or sequences of scenes) generated during the execution of a test case (test case evaluation data).
A metric result is calculated at a certain point in time and consists of a number and a unit.
The calculated metric results are evaluated according to the corresponding evaluation criteria.

{\color{alizarin}\Pifont{pzd}{\char227}} 
\textbf{Test case evaluation data}:
Test case evaluation data are the data generated during the execution of a test case with the intention of calculating metric results.
These data can be recorded for a later test case evaluation or evaluated simultaneously with the execution of the test case.

{\color{alizarin}\Pifont{pzd}{\char227}} 
\textbf{Threshold value}:
A threshold value is a fixed number (with a unit) used to test the compliance of the calculated metric results according to the evaluation criterion.
Therefore, only a statement can indicate whether the evaluation criterion is fulfilled or not.
 
{\color{alizarin}\Pifont{pzd}{\char227}} 
\textbf{Evaluation scale}:
An evaluation scale is a scale used to evaluate the adherence of the calculated metric results with this scale according to the evaluation criterion.
Therefore, it is also possible to make a statement about how well the evaluation criterion is fulfilled.

{\color{alizarin}\Pifont{pzd}{\char227}} 
\textbf{Application period}:
An application period defines the periods in which the corresponding evaluation criterion is applied. 
The application period is defined by one or more conditions linked with AND and/or OR operators. 
When the linked conditions are fulfilled, the application of the evaluation criterion begins. 
Its application continues until the linked conditions are no longer fulfilled, a specified time has elapsed, or a specified event has occurred.

\subsubsection{Test Configuration Phase} \label{ssec_TestConfiguration}

The test configuration phase follows the test design and implementation phase.
In this test phase, one or more test bench configurations are created.
These test bench configurations are either directly specified in the test procedure specification or created manually or automatically based on the test specification.
A test bench configuration is a composition of one or more elements (e.g.,~several simulation models) and their specific parameter values, which are provided at a test bench and are necessary to execute one or more test cases. 
Thus, depending on the elements provided at a test bench, one test bench provides one or more test bench configurations.
We consider a differentiation between the terms \emph{test bench} and \emph{test bench configuration} meaningful since in practice, a test bench is not rebuilt for each test case; rather, it exists or is built to execute several test cases.

\Figure[t!](topskip=0pt, botskip=0pt, midskip=0pt)[]{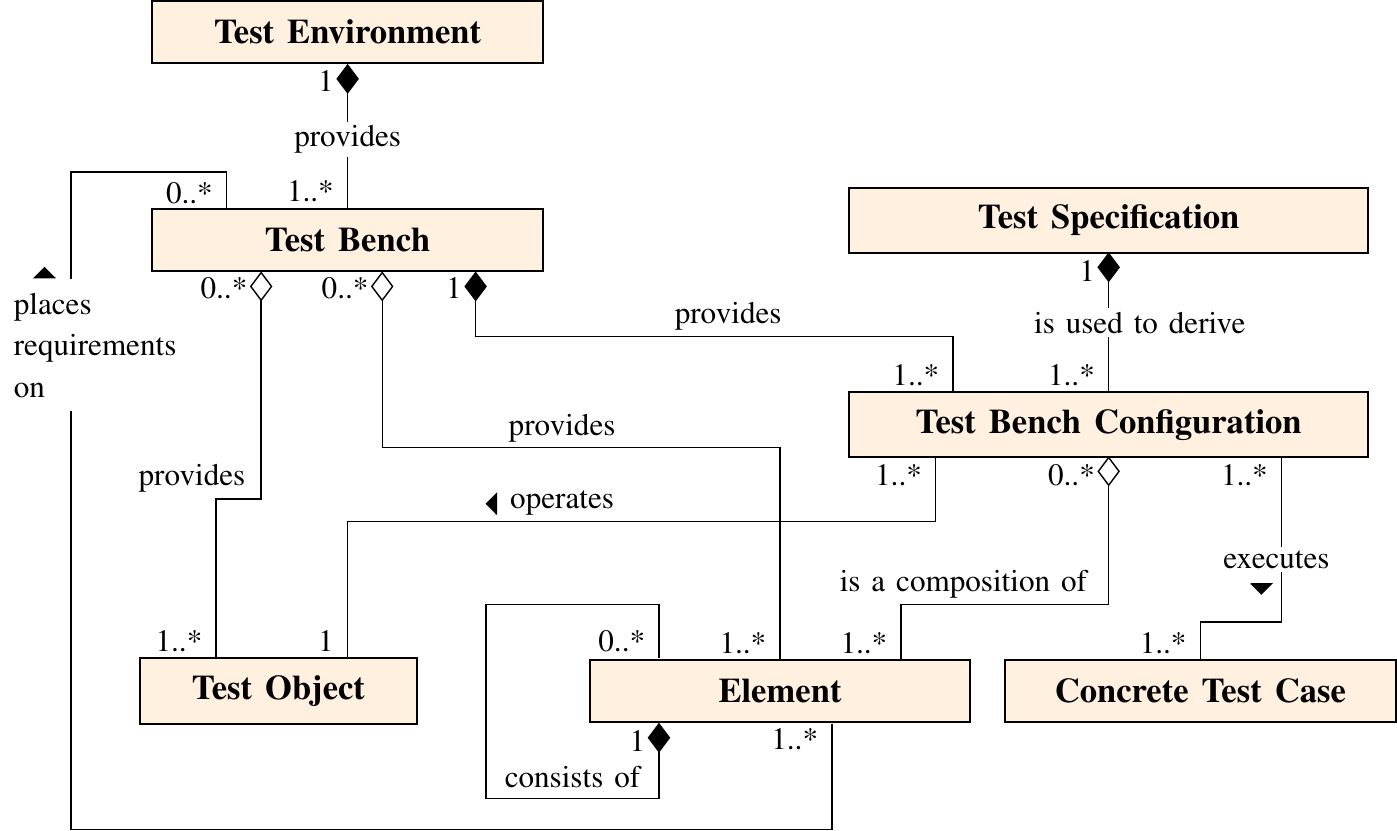}
{Terms proposed for the basic vocabulary related to the test configuration phase visualized as a UML diagram.\label{fig_TestConfiguration_UML}}

Fig.~\ref{fig_TestConfiguration_UML} visualizes the terms related to the test configuration phase that we consider particularly relevant to an overview of scenario-based development and test approaches for automated vehicles and their relationships as a UML diagram. 
In the following paragraphs, we describe the terms illustrated in this UML diagram, which have not been described thus far.

{\color{alizarin}\Pifont{pzd}{\char227}} 
\textbf{Test environment}:
The test environment consists of ``facilities, hardware, software, firmware, procedures, and documentation intended for or used to perform testing [...]\footnote{``of software'' was used in the original text; an explanation of its deletion is provided in the text above.}.''~\cite{ISO29119_2013}
The test environment provides one or more test benches.

{\color{alizarin}\Pifont{pzd}{\char227}} 
\textbf{Test bench}:
A test bench is a ``technical device'' (consisting of hardware and software) that provides test objects and elements intended to execute test cases.
Depending on the elements provided at a test bench, a test bench provides one or more test bench configurations due to the different compositions of the elements that can be used to execute test cases.

{\color{alizarin}\Pifont{pzd}{\char227}} 
\textbf{Test bench configuration}:
A test bench configuration is a composition of one or more elements and their specific parameter values, which are provided at a test bench and are necessary to execute one or more test cases.

{\color{alizarin}\Pifont{pzd}{\char227}} 
\textbf{Element (of a test bench)}:
An element (of a test bench) is a piece of hardware and/or software that is provided at a test bench and can be used to execute a test case; examples include a specific simulation model, a balloon vehicle, or a driving robot. 
An element can consist of subelements.

\subsubsection{Test Execution Phase}   \label{ssec_TestExecution}

The test execution phase follows the test configuration phase.
In this test phase, one or more test cases are executed as specified in the test procedure specification using one or more test bench configurations.
During the execution of the test case, the test bench configuration~--~more precisely, the elements contained in the test bench configuration~--~generate test object input values that stimulate the test object based on the scene at time~$t_s$.
The test object processes these test object input values and generates the test object output values at the next point in time~($t_{s+1}$).
As the scenario progresses, the test bench configuration generates the scene at the next point in time~($t_{s+1}$) based on the scene at time~$t_s$ and the behavior of all objects involved.
For this purpose, the test bench configuration can read zero or more than zero test object output values and consider them when generating the scene at time~$t_{s+1}$ (closed loop) or not consider them (open loop).
During the execution of the test case, the test bench configuration generates data that can be used to evaluate the test case.

Fig.~\ref{fig_TestCaseExecution_UML} visualizes the terms related to the test execution phase that we consider particularly relevant to an overview of scenario-based development and test approaches for automated vehicles and their relationships as a UML diagram. 
In the following paragraphs, we describe the terms illustrated in this UML diagram, which have not been described thus far.

\Figure[t!](topskip=0pt, botskip=0pt, midskip=0pt)[]{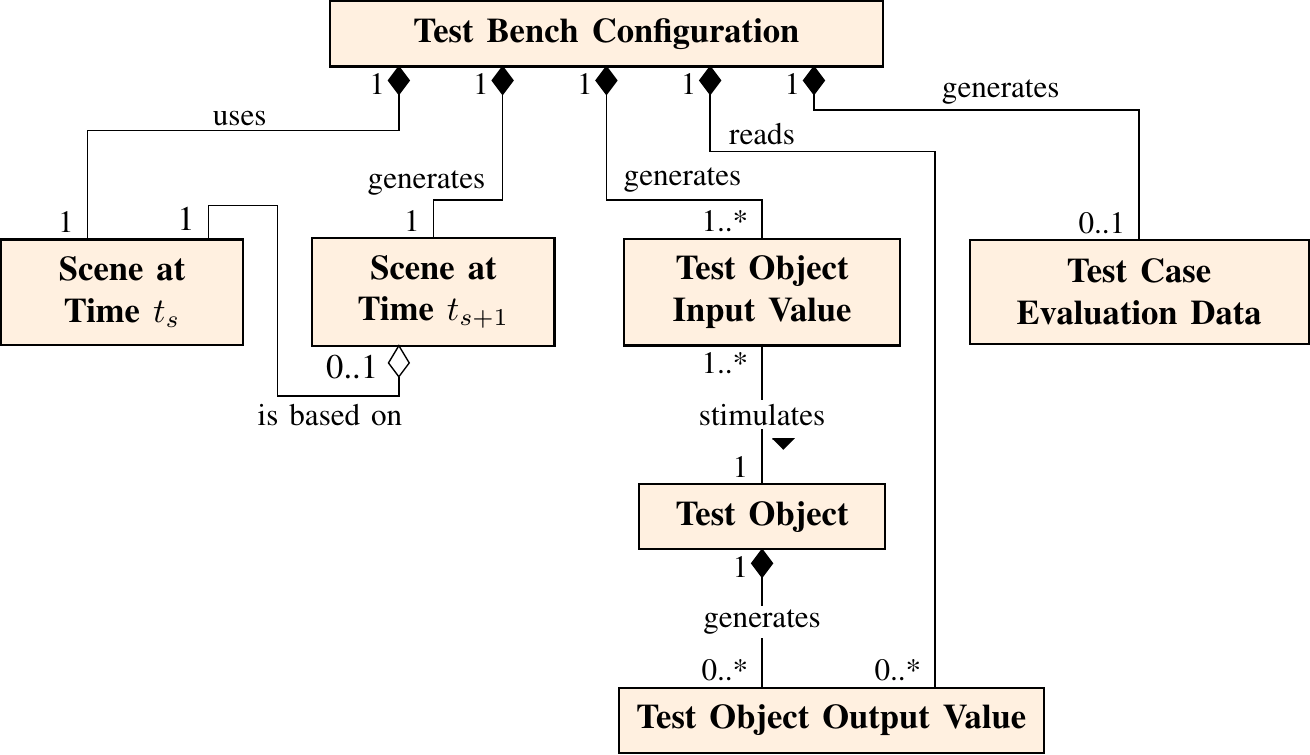}
{Terms proposed for the basic vocabulary related to the test execution phase visualized as a UML diagram.\label{fig_TestCaseExecution_UML}}

{\color{alizarin}\Pifont{pzd}{\char227}} 
\textbf{Test object input value}:
A test object input value is a value that stimulates the test object at a specific point in time.
Test object input values are generated based on the scene at the corresponding point in time.

{\color{alizarin}\Pifont{pzd}{\char227}} 
\textbf{Scene at a specific point in time}:
The scene at a specific point in time is a snapshot of the environment at the corresponding point in time, including the movable objects, the scenery, all actors’ and observers’ self-representations, and the relationships among these entities.

{\color{alizarin}\Pifont{pzd}{\char227}} 
\textbf{Test object output value}:
A test object output value is a value generated by the test object at a specific point in time.

\subsubsection{Test Evaluation and Reporting Phase} \label{ssec_TestEvaluation}

The test evaluation and reporting phase follows the test execution phase.
In this test phase, the test case evaluation data are evaluated according to the evaluation criteria, and a test report is generated.
For this purpose, one or more metric results are calculated based on the test case evaluation data according to one or more evaluation metrics.
The calculated metric results are evaluated according to the evaluation criteria that have threshold values or evaluation scales.
This evaluation can be performed simultaneously with or after the execution of the test case based on the recorded data.
A test report is generated based on all evaluations performed.
Based on the evaluation of all test cases required for, as an example, the release of an automated vehicle, an overall evaluation of the vehicle can be performed.
Based on this overall evaluation attributes such as statistical proof of safety can be provided, which in turn can be used as a basis for the vehicle's release.

Fig.~\ref{fig_TestEvaluation_UML} visualizes the terms related to the test evaluation and reporting phase that we consider particularly relevant to an overview of scenario-based development and test approaches for automated vehicles and their relationships as a UML diagram. 
In the following paragraphs, we describe the terms illustrated in this UML diagram, which have not been described thus far.

\Figure[t!](topskip=0pt, botskip=0pt, midskip=0pt)[]{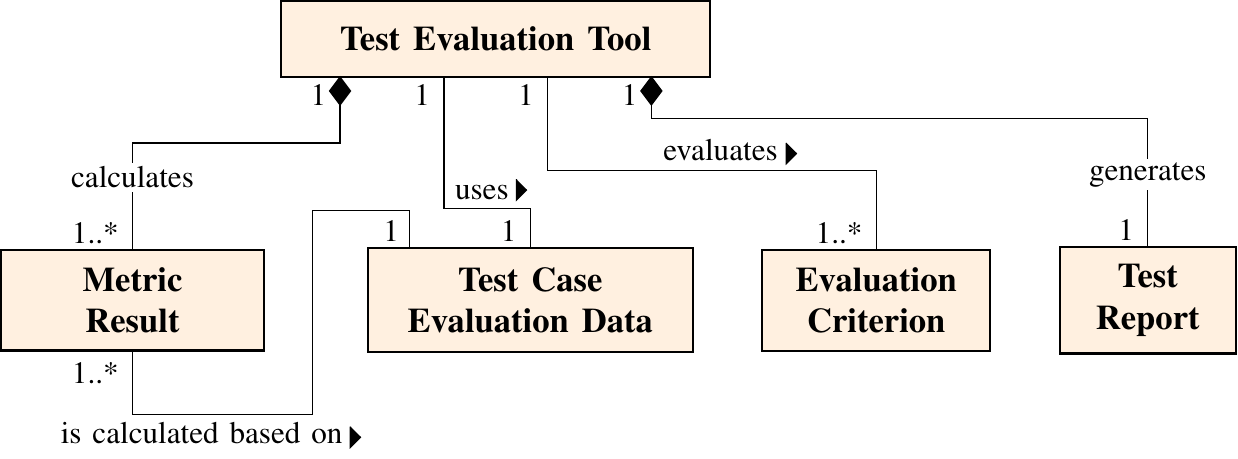}
{Terms proposed for the basic vocabulary related to the test evaluation and reporting phase visualized as a UML diagram.\label{fig_TestEvaluation_UML}}

{\color{alizarin}\Pifont{pzd}{\char227}} 
\textbf{Test evaluation tool}:
A test evaluation tool calculates one or more metric results based on test case evaluation data according to one or more evaluation metrics within the application period.
This tool evaluates these calculated metric results according to the evaluation criteria that contain threshold values or evaluation scales.
This tool summarizes the performed evaluations of one or more executed test cases into an overall test result and generates a test report.

{\color{alizarin}\Pifont{pzd}{\char227}} 
\textbf{Test report}:
A test report is a document that describes the evaluations performed and the overall test result.

\section{Application of the proposed basic vocabulary for scenario-based development \\and test approaches for automated vehicles} \label{sec_application}

In this section, the terms introduced in Section~\ref{sec_terminology} are applied, explained, and distinguished from each other based on the development and testing of an ACC system. 
Examples are provided for each term proposed for the basic vocabulary to clarify and distinguish the terms. 
Clarification based on an ACC system seems sensible, as many people are familiar with such a system. 
However, the examples provided are intended not to develop an ACC system but to clarify the respective terms. 
That is why the examples provided are deliberately kept simple. 
To indicate the positions where the respective terms are explained by way of example, these terms are printed in bold text.

\subsection{Scenario Derivation and Usage} \label{ssec_ScenarioDerivationExample}

In this subsection, terms related to the scenario derivation and usage, which are introduced in Section~\ref{ssec_scenarioDerivation}, are explained and distinguished from each other based on the development and testing of an ACC system. 
In this example, one functional, one logical, and one concrete scenario are presented.

Fig.~\ref{fig_scenario_UML} visualizes the proposed terms related to a scenario and their relationships as a UML diagram.
Fig.~\ref{fig_scenario_levels_UML} visualizes the proposed terms related to the different abstraction levels of scenarios and their relationships as a UML diagram.

A possible \textbf{functional scenario}, which is described linguistically, is that the ego vehicle (vehicle under consideration that is equipped with the ACC system) drives at a certain initial speed on the right-hand lane of a two-lane highway with one curve and without other road users in the summer. 
At a specific point in time, the ACC system is activated by setting a maximum desired speed for the ego vehicle (called the set speed).
From this point on, the ACC system automatically controls the ego vehicle's speed.

As the development process progresses, the derived functional scenarios are detailed by logical scenarios. 
From the functional scenario described above, the \textbf{logical scenario} described below can be derived. 
It is referred to as ``Logical Scenario \textit{SpeedControl}'' in this publication. 
For clarity and simplicity, we do not list all of the parameters needed to describe this logical scenario completely; rather, we provide only the excerpt that is necessary to illustrate the proposed terms.
This excerpt is shown graphically in Fig.~\ref{fig_LogicalScenario} and is described below.
The \textbf{scenery} in this logical scenario consists of a two-lane highway, among other aspects.
Each lane has a width of \SIrange{2.5}{3.9}{\meter} and a curvature of ${0}$\,\SI{}{\meter\tothe{-1}} to ${1\text{E}-3}$\,\SI{}{\meter\tothe{-1}}.
The environmental temperature is between \SIrange[range-phrase=~and~]{5}{35}{\celsius}.
The ego vehicle is a \textbf{movable object}.
The initial speed of the ego vehicle is between \SIrange[range-phrase=~and~]{80}{160}{\kilo \meter \per \hour}.
There are no other movable objects in this scenario.
The \textbf{self-representation} of the ego vehicle includes the functional readiness of the ACC system according to its item definition.
One to \SI{10}{\second} after the scenario starts, the \textbf{event}, activating the ACC system, occurs by setting a set speed between \SIrange[range-phrase=~and~]{60}{200}{\kilo \meter \per \hour}.
An individual \textbf{goal} of the system is to maintain the set speed.
An individual \textbf{goal} of the driver is to reach the destination, Brunswick. 
An individual \textbf{value} of the driver is his or her need for safety, which depends, among other factors, on the desired distance to a vehicle in front of the ego vehicle (small, medium, or large).

As the development process progresses, the derived logical scenarios are concretized by concrete scenarios. 
From the logical scenario described above, the concrete scenario described below can be derived.
For this purpose, each parameter from the logical scenario is assigned a fixed value.
Based on the parameters used to describe the logical scenario mentioned above, the \textbf{concrete scenario} described below can be derived.
It is referred to as ``Concrete Scenario \textit{SpeedControl}'' in this publication.
The excerpt necessary to illustrate the proposed terms is shown graphically in Fig.~\ref{fig_ConcreteScenario}.
The \textbf{scenery} in this concrete scenario consists of a two-lane highway, among other aspects.
Each lane has a width of \SI{3.75}{\meter} and a curvature of ${5\text{E}-4}$\,\SI[parse-numbers = false]{}{\meter\tothe{-1}}.
The environmental temperature is \SI{20}{\celsius}.
The ego vehicle is a \textbf{movable object}.
The initial speed of the ego vehicle is \SI{150}{\kilo \meter \per \hour}.
There are no other movable objects in this scenario.
The \textbf{self-representation} of the ego vehicle includes the functional readiness of the ACC system according to its item definition.
Two~seconds after the scenario starts, the \textbf{event}, activating the ACC system, occurs by setting a set speed of \SI{120}{\kilo \meter \per \hour}.
An individual \textbf{goal} of the system is to maintain the set speed.
An individual \textbf{goal} of the driver is to reach the destination, Brunswick. 
An individual \textbf{value} of the driver is his or her need for safety, which depends, among other factors, on the desired distance to a vehicle in front of the ego vehicle.
In this concrete scenario, the driver has a high need for safety; therefore, he or she chooses a large distance to vehicles in front of the ego vehicle.

Through a \textbf{real-world test drive} on a highway, data can be recorded. 
By analyzing these data, concrete scenarios can be identified and assigned to existing or to-be-created logical scenarios. 
In this example, the width of the lanes (in this recording, \SI{3.4}{\meter}) and the curvature (in this recording, ${5\text{E}-3}$\,\SI[parse-numbers = false]{}{\meter\tothe{-1}}) can be extracted. 
If the other parameter values also correspond to the parameter values in the logical scenario mentioned above, this identified concrete scenario can be assigned to this logical scenario.

\subsection{Concept, Design, and Implementation Phase} \label{ssec_DesignImplementationExample}

In this subsection, terms related to the concept, design, and implementation phase, which are introduced in Section~\ref{ssec_DesignImplementation}, are explained and distinguished from each other based on the development and testing of an ACC system. 
Fig.~\ref{fig_ProductDevelopment_UML} visualizes the proposed terms related to this phase and their relationships as a UML diagram.

The \textbf{vehicle function} represents the functional behavior of the vehicle.
Based on the ISO~15622 standard~\cite{ISO15622_2010}, the vehicle function being developed is its ability to control the ego vehicle's speed automatically either to maintain a clearance to a vehicle in front of the ego vehicle (following control) or to maintain the set speed (speed control), whichever is lower.
For this purpose, the engine, the power train, and the brakes are controlled. 

In this example, the vehicle function described above will be implemented by one \textbf{item}, which is the ACC system to be developed.
The ACC system is specified by an \textbf{item definition}.
The item definition specifies, among other things, the functionality implemented by the ACC system.
Since the vehicle function will be implemented by only one item in this example, the \textbf{functionality} specified in the item definition corresponds to the vehicle function described above.
To describe (a part of) this functionality, two (abstracted) \textbf{functional scenarios} are listed below.
These scenarios describe the functional behavior of the ego vehicle at the vehicle level.

\begin{itemize}
	\item When driving in speed control mode, if the ego vehicle's current speed is lower than the set speed, the ACC system will accelerate the ego vehicle to the set speed. 
	\item When driving in speed control mode, if the ego vehicle's current speed is higher than the set speed, the ACC system will decelerate the ego vehicle to the set speed. 
\end{itemize}

\Figure[t!](topskip=0pt, botskip=0pt, midskip=0pt)[]{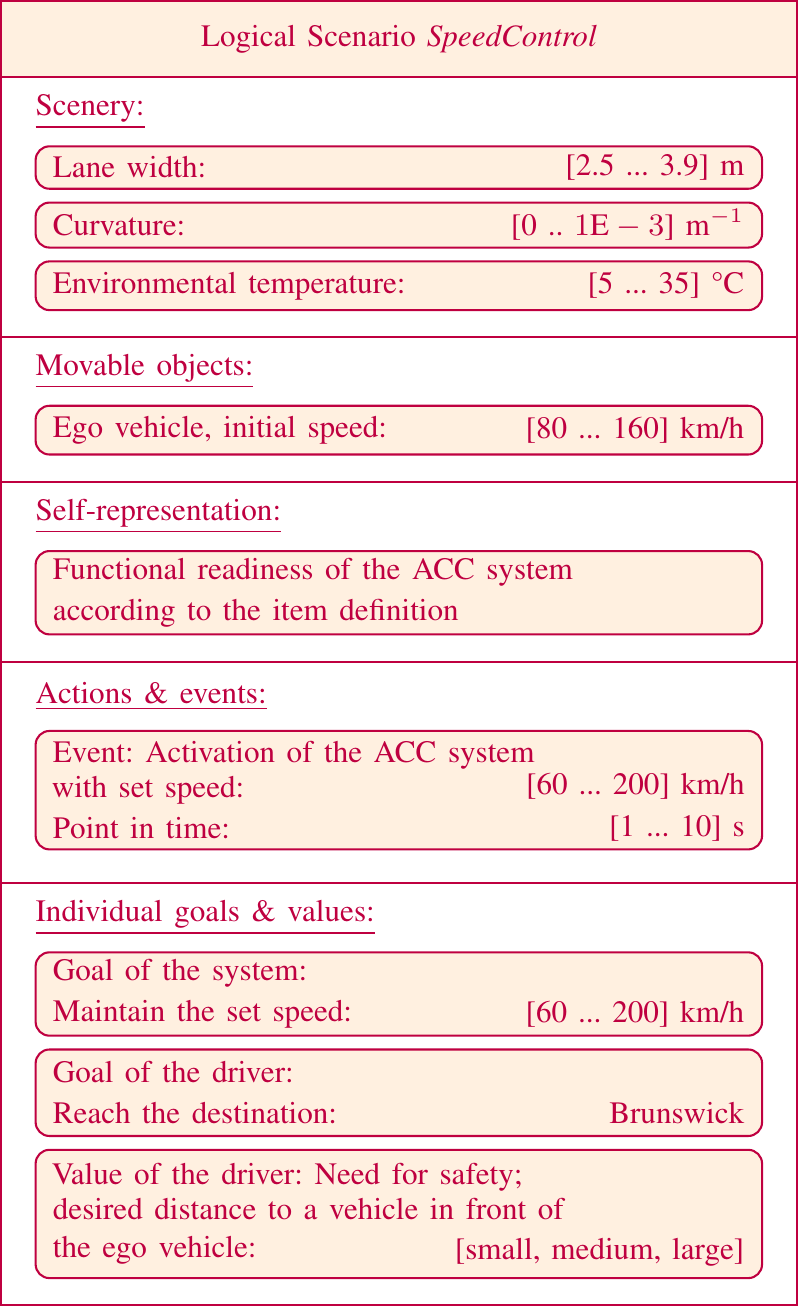}
{Excerpt from the parameter space of the considered logical scenario, which illustrates the proposed terms related to a logical scenario.\label{fig_LogicalScenario}}

\Figure[t!](topskip=0pt, botskip=0pt, midskip=0pt)[]{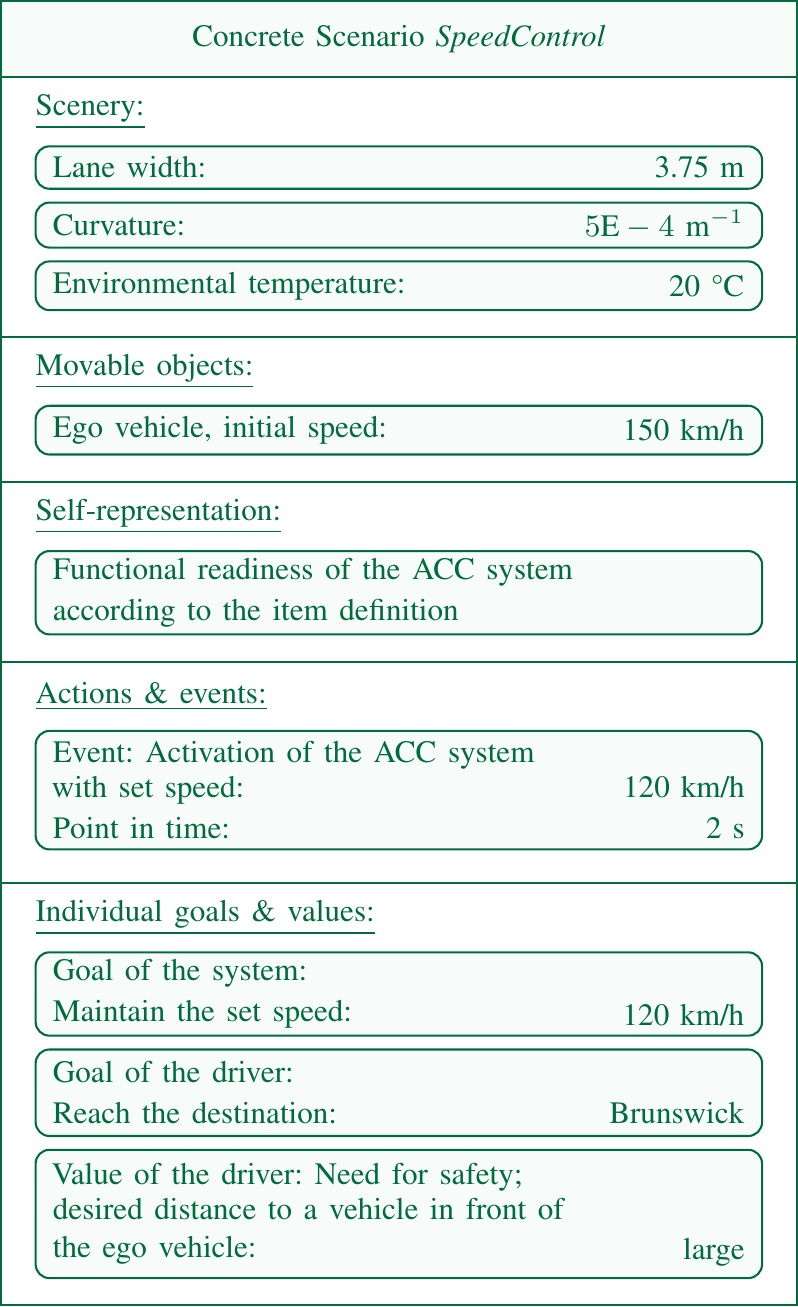}
{Excerpt from the parameter values of the considered concrete scenario, which illustrates the proposed terms related to a concrete scenario.\label{fig_ConcreteScenario}}

Additionally, the item definition can be used as a basis for deriving the ACC system's requirements.
As an example, we present two requirements based on the scenarios listed above and on the ISO~15622 standard~\cite{ISO15622_2010}. 
They are not intended to represent the ACC system entirely but only two basic requirements, which are used in this publication to explain the proposed terms.

\begin{itemize}
	\item \textbf{First requirement}: When driving in speed control mode, the ACC system automatically controls the ego vehicle's speed to maintain the set speed (desired travel speed, set either by the driver or by a control system) with a maximum deviation of $\SI{-5}{\percent}$ from the set speed.

	\item \textbf{Second requirement}: The average automatic deceleration of the ego vehicle by the ACC system must not exceed \SI{3.5}{\meter \per \second\squared} in a \SI{2}{\second} interval. 

\end{itemize}

\Figure[t!](topskip=0pt, botskip=0pt, midskip=0pt)[]{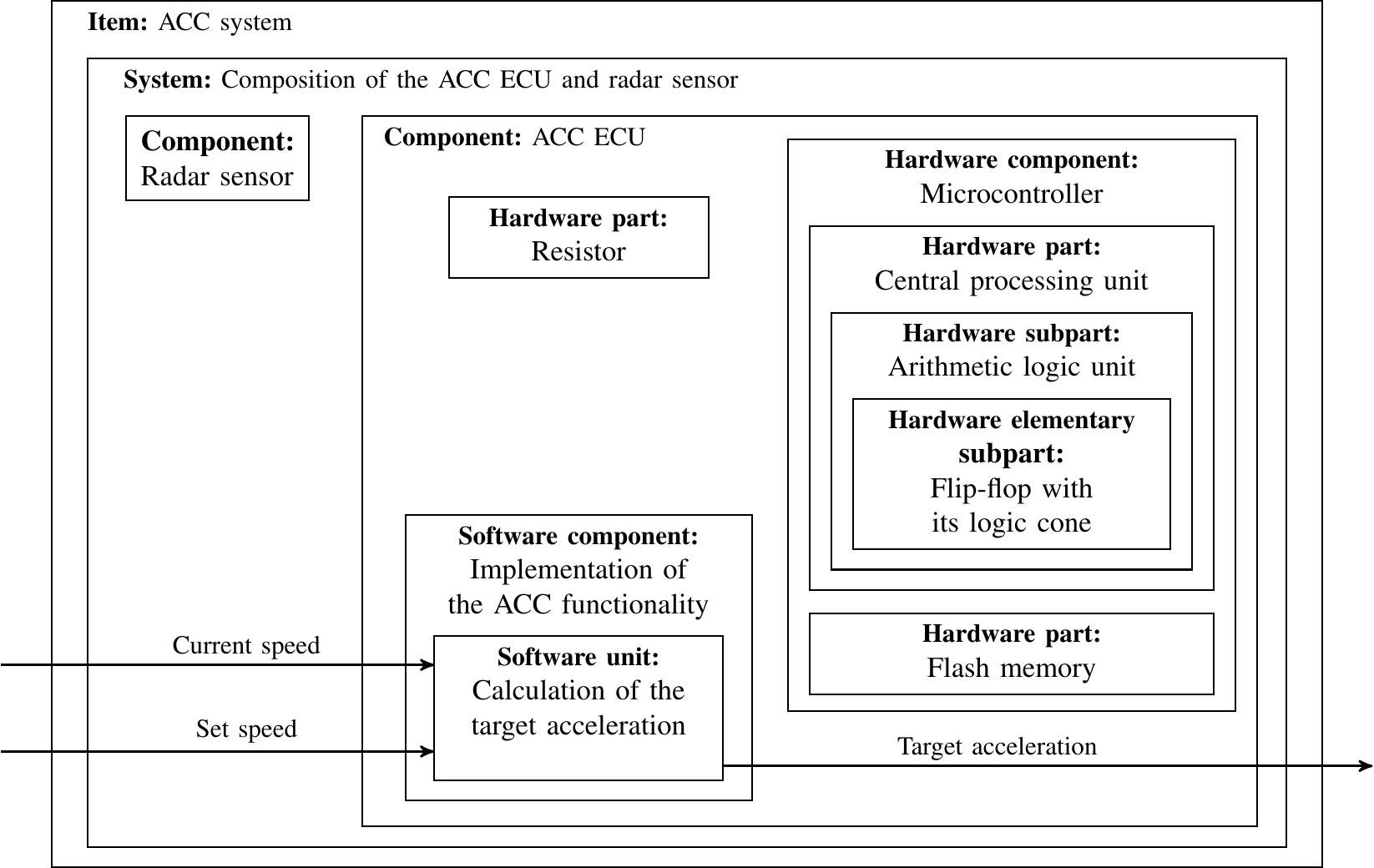}
{Simplified example subdivision of the ACC system into different elements, which illustrates the terms system, component, hardware part, and software unit (only a subset of the elements necessary to explain the proposed terms are provided as examples; ECU: electronic control unit).\label{fig_DevisionSystem}}

An item consists of several elements.
Fig.~\ref{fig_DevisionSystem} illustrates an excerpt from a simplified example subdivision of the item considered in this example, that is, the ACC system, into different elements to illustrate the terms mentioned above. 
The elements are partially based on the elements provided as examples in the ISO~26262 standard~\cite{ISO26262_2018}.
For clarity, only a subset of the elements necessary to explain the proposed terms is provided as an example.
In reality, considerably more elements are required. 
For simplicity, an ACC \textbf{system} is considered a composition of the ACC electronic control unit (ECU) and the radar sensor. 
A \textbf{component} of this system is the radar sensor.
Another \textbf{component} of this system is the ACC ECU, which consists of hardware components, hardware parts, software components, and software units. 
A \textbf{hardware component} is the deployed microcontroller.
\textbf{Hardware parts} belonging to this microcontroller include the central processing unit and the flash memory.
A \textbf{hardware subpart} of the microcontroller is its arithmetic logic unit, which includes a flip-flop with its logic cone as a \textbf{hardware elementary subpart}.
Another \textbf{hardware part} of the ACC ECU is a resistor. 
A \textbf{software component} of the ACC ECU is the software implementation of the ACC functionality.
A \textbf{software unit} of this software component is the software used to calculate the target acceleration of the ego vehicle. 
For this purpose, this software unit receives the ego vehicle's current speed and the set speed as input values. 
Based on these input values, this software unit calculates the target acceleration of the ego vehicle and outputs the calculated value.

\subsection{Test Process}

In this subsection, terms related to the test process, which are introduced in Section~\ref{ssec_TestProcess}, are explained and distinguished from each other based on the development and testing of an ACC system.

\subsubsection{Test Planning Phase} \label{ssec_TestPlanningExample}

In this subsection, terms related to the test planning phase, which are introduced in Section~\ref{ssec_TestPlanning}, are explained and distinguished from each other based on the development and testing of an ACC system. 
Fig.~\ref{fig_TestPlanning_UML} visualizes the proposed terms related to this test phase and their relationships as a UML diagram.
This publication does not aim to provide a detailed description of the documents created during this test phase. 
Therefore, these documents are described only to the extent necessary to explain the terms that depend on them.
For this reason, they are described in a simplified manner and do not claim to be comprehensive.

The \textbf{project plan} describes the overall project for developing the ACC system.
The \textbf{organizational test strategy} describes the testing of all projects within an organization.
A detailed description of these documents is not the aim of this publication; therefore, we do not discuss them in detail.

The \textbf{project test plan} is derived from the project plan and describes the overall testing strategy and the test processes to be used.
In this example, one testing context is to verify the compliance of the implemented ACC system with the specified requirements at the following test levels: the vehicle level, the system level, the component level, and the unit level.
Therefore, the \textbf{overall test process}, which is described in this project test plan, is divided into the following \textbf{test sub-processes}: a vehicle test sub-process, a system test sub-process, a component test sub-process, and a unit test sub-process.

Every test sub-process is described in detail in a \textbf{sub-process test plan}, which is derived from the project test plan.
In summary, there are four sub-process test plans in this example: a vehicle test plan, a system test plan, a component test plan, and a unit test plan.

Continuing with this example, the component test plan is considered.
This \textbf{test plan} describes the test objective, identifies the test object and the test scope, and includes the test strategy for the component test sub-process.
The \textbf{test object} is the component named the ``ACC ECU'' (version~1.0), which is described in Section~\ref{ssec_DesignImplementationExample}.
The \textbf{test objective} in this example is to verify the compliance of this component with the specified requirements.
The \textbf{test scope} summarizes the features of the test object to be tested, including the two requirements listed in Section~\ref{ssec_DesignImplementationExample}.

The test strategy for the component test sub-process is derived based on the organizational test strategy.
As described above, the \textbf{test strategy} describes ``the test practices used; the test sub-processes to be implemented; the retesting and regression testing to be employed; the test design techniques and corresponding test completion criteria to be used; test data; test environment and testing tool requirements; and expectations for test deliverables.''~\cite{ISO29119_2013}
A detailed description of a test strategy is not the aim of this publication; therefore, we do not introduce any details.

According to the \textbf{test design technique}, the test cases are derived based on the derived concrete scenarios and the requirements. 
Additionally, the test cases are specified in such a way that they can be executed on the available hardware-in-the-loop test bench.

\subsubsection{Test Design and Implementation Phase}
\label{ssec_TestSpecificationExample}

In this subsection, terms related to the test design and implementation phase, which are introduced in Section~\ref{ssec_TestSpecification}, are explained and distinguished from each other based on the development and testing of an ACC system. 
Fig.~\ref{fig_TestSpecification_UML} visualizes the proposed terms related to this test phase and their relationships as a UML diagram.

The test specification is created based on the specified scenarios, the component test plan, and the requirements.
To create the test specification, three elements are used: the concrete scenario ``Concrete Scenario \textit{SpeedControl}'' introduced in Section~\ref{ssec_ScenarioDerivationExample}, the two requirements listed in Section~\ref{ssec_DesignImplementationExample}, and the component test plan described above.
The \textbf{test specification} consists of the test design specification, the test case specification, and the test procedure specification for the considered test object. 
These terms are explained in the following paragraphs.

The test design specification is a document that specifies the features to be tested and their corresponding test conditions.
In this example, the \textbf{test design specification} demands that all requirements be verified. 
For this purpose, the available hardware-in-the-loop test bench should be used. 
Among other things, this means that the concrete scenario must be specified using the OpenDRIVE and OpenSCENARIO formats.
Additionally, the test design specification specifies that the test cases for the considered test object will be derived based on the derived concrete scenarios and the requirements. 
Again, a detailed description of the test design specification is not the aim of this publication; therefore, we do not introduce any details.

The \textbf{test case specification} is a document that specifies the test cases for the considered test object. 
Their derivation is described in the test design specification.
In this example, one \textbf{test case} is derived based on the concrete scenario ``Concrete Scenario \textit{SpeedControl}'' introduced in Section~\ref{ssec_ScenarioDerivationExample}, which is combined with the following two evaluation criteria derived from the two requirements listed in Section~\ref{ssec_DesignImplementationExample}.

Based on the first requirement listed in Section~\ref{ssec_DesignImplementationExample}, the \textbf{first evaluation criterion} is derived for this test case; it is shown graphically in Fig.~\ref{fig_firstConcreteEvaluationCriterion}.
This evaluation criterion states that the ACC system should automatically control the ego vehicle's speed to maintain its speed with a maximum deviation of $\SI{-5}{\percent}$ from the set speed~$v_\text{set}$.
Thus, two conditions can be defined for the \textbf{application period} of this evaluation criterion.
The first condition states that the ACC system must be active. 
The second condition states that the ego vehicle must reach the set speed~$v_\text{set}$ for the first time.
Both conditions are linked with an AND operator.
Eq.~\ref{eq_application_period_1} defines the corresponding application period.
The application of this evaluation criterion begins when these linked conditions are fulfilled and lasts until the end of the scenario (after activating the ACC system, it is active until the end of the scenario).
The \textbf{mathematical formula} used to calculate the \textbf{metric results} consists of recording the ego vehicle's speed in each scene when the application period is active (see Eq.~(\ref{eq_evaluation_function_1})).
This mathematical formula belongs to the \textbf{evaluation metric} referred to by the \textbf{evaluation metric name} ``Ego\_speed''.
An \textbf{evaluation scale} is defined for this evaluation criterion so that a larger deviation in the ego vehicle's speed from the set speed is rated worse than a smaller deviation. 
If the set speed is met, the evaluation criterion is $\SI{100}{\percent}$ fulfilled. 
If the ego vehicle's current speed deviates by more than $\SI{-5}{\percent}$ from the set speed or if the ego vehicle's current speed exceeds the set speed, the evaluation criterion is $\SI{0}{\percent}$ fulfilled.
In this example, a linear evaluation lies between these limits (see Fig.~\ref{fig_firstConcreteEvaluationCriterion}).

\begin{align}
	\label{eq_application_period_1}
	&\text{AP}_{\text{EC1}} := \text{ACC(active)~} \&\& \text{~(}v_{\text{Ego}} == v_\text{set}\text{)}^1  \\
	& \text{where:} \notag \\
	&\text{AP}_{\text{EC1}}: \text{application period of the first evaluation criterion} \notag \\
	&\text{ACC(active)}: \text{ACC system is active} \notag \\
	&v_\text{Ego}: \text{ego vehicle's speed} \notag \\
	&v_\text{set}: \text{set speed} \notag \\
	&^1: \text{must be fulfilled once} \notag 
\end{align}

\begin{align}
	\label{eq_evaluation_function_1}
	&\text{MR}_\text{v,ego}(t_\text{s}) = v_{\text{Ego}} (\text{scene}(t_\text{s})) \\
	& \text{where:} \notag \\
	&t_\text{s}: \text{considered point in time} \notag \\
	&\text{MR}_\text{v,ego}(t_\text{s}): \text{metric result calculated with the evaluation} \notag \\ 
	&\hspace{1.9cm}\text{metric named ``Ego\_speed'' at time $t_\text{s}$} \notag \\
	&\text{scene}(t_\text{s}): \text{scene at time $t_\text{s}$} \notag \\	
	&v_\text{Ego}(\text{scene}(t_\text{s})): \text{ego vehicle's speed in scene at time $t_\text{s}$} \notag 
\end{align}

\Figure[t!](topskip=0pt, botskip=0pt, midskip=0pt)[]{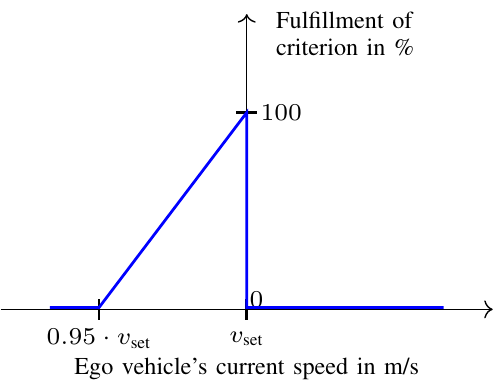}
{Visualization of the first evaluation criterion, $v_\text{set}:$ set speed.\label{fig_firstConcreteEvaluationCriterion}}

Based on the second requirement listed in Section~\ref{ssec_DesignImplementationExample}, the \textbf{second evaluation criterion} is derived for this test case; it is shown graphically in Fig.~\ref{fig_secondConcreteEvaluationCriterion}.
This criterion states that the average automatic deceleration in a \SI{2}{\second} interval by the ACC system must not exceed \SI{3,5}{\meter \per \second\squared}.
Thus, one condition can be defined for the evaluation criterion's \textbf{application period}, which states that the ACC system has to be active.
Eq.~\ref{eq_application_period_2} defines the corresponding application period.
The application of this evaluation criterion begins when this condition is fulfilled and lasts until the end of the scenario (after activating the ACC system, it is active until the end of the scenario).
The \textbf{mathematical formula} used to calculate the \textbf{metric results} consists of recording the ego vehicle's deceleration in sequences of scenes covering time intervals of \SI{2}{\second} each and calculating the average deceleration within these intervals when the application period is active (see Eq.~(\ref{eq_evaluation_function_2})).
This mathematical formula belongs to the \textbf{evaluation metric} referred to by the \textbf{evaluation metric name} ``Average\_ego\_deceleration''.
A \textbf{threshold value} of \SI{3,5}{\meter \per \second\squared} is defined for this evaluation criterion.
If the average automatic deceleration is below this threshold value in the corresponding \SI{2}{\second} interval, this evaluation criterion is $\SI{100}{\percent}$ fulfilled.
If it is above this threshold, this evaluation criterion is $\SI{0}{\percent}$ fulfilled (see Fig.~\ref{fig_secondConcreteEvaluationCriterion}).

\begin{align}
	\label{eq_application_period_2}
	&\text{AP}_{\text{EC2}} := \text{ACC(active)~} \\ 
	& \text{where:} \notag \\
	&\text{AP}_{\text{EC2}}: \text{application period of the second evaluation criterion} \notag \\
	&\text{ACC(active)}: \text{ACC system is active} \notag 
\end{align}

\begin{align}
	\label{eq_evaluation_function_2}
	&\text{MR}_\text{${\bar a}$,ego}(t_\text{s}) = \frac{1}{n} \cdot \sum_{s=\text{scene}(t_\text{s-2\,sec})}^{s=\text{scene}(t_\text{s})}  a_{\text{Ego}}(s) \\
	& \text{where:}  \notag \\
	&t_\text{s}: \text{considered point in time} \notag \\
	&\text{MR}_\text{${\bar a}$,ego}(t_\text{s}): \text{metric result calculated with the evaluation} \notag \\
	&\hspace{2cm}\text{metric named ``Average\_ego\_deceleration''} \notag  \\
	&\hspace{2cm}\text{at time $t_\text{s}$} \notag  \\
	&\text{scene}(t_\text{s}): \text{scene at time $t_\text{s}$} \notag \\	
	&\text{scene}(t_\text{s-2\,sec}): \text{scene at time $t_\text{s}$ - 2\,\text{seconds}} \notag \\	
	&a_{\text{Ego}}(s): \text{ego vehicle's deceleration in the considered scene} \notag \\ 
	&n: \text{number of scenes covering 2\,seconds} \notag 
\end{align}

\Figure[t!](topskip=0pt, botskip=0pt, midskip=0pt)[]{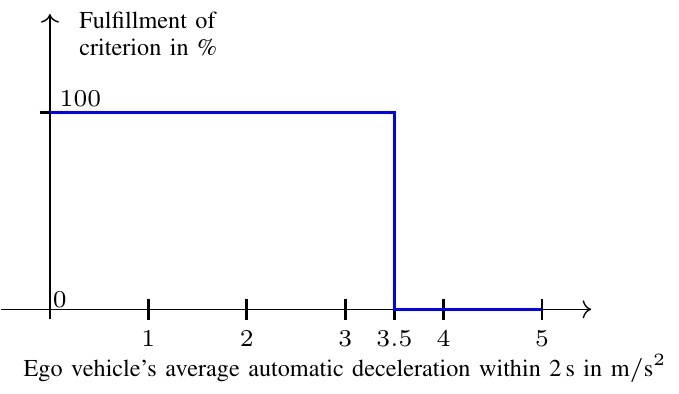}
{Visualization of the second evaluation criterion.\label{fig_secondConcreteEvaluationCriterion}}

In this publication, combining the concrete scenario ``Concrete Scenario \textit{SpeedControl}'' and the two evaluation criteria introduced above results in a test case called ``Test Case \textit{SpeedControl}''.
This test case is specified by the test case specification.

The \textbf{test procedure specification} is a document that specifies one or more test procedures, which are collections of test cases to be executed for the considered test objective.
Since there is only one test case in this example, there is only one \textbf{test procedure}, including this test case.
This test procedure does not include any further evaluation criterion.
Additionally, it is specified that this test procedure shall be executed using the hardware-in-the-loop test bench.
A detailed description of the test procedure specification is not the aim of this publication; therefore, we do not introduce any details.

\subsubsection{Test Configuration Phase}
\label{ssec_TestConfigurationExample}

In this subsection, terms related to the test configuration phase, which are introduced in Section~\ref{ssec_TestConfiguration}, are explained and distinguished from each other based on the development and testing of an ACC system. 
Fig.~\ref{fig_TestConfiguration_UML} visualizes the proposed terms related to this test phase and their relationships as a UML diagram.

In this example, the \textbf{test environment} consists of all facilities, hardware, software, firmware, procedures, and documentation used to perform testing.
This test environment provides, among other things, the hardware-in-the-loop test bench and all the components and elements mentioned below.
A more detailed description of this test environment is not the aim of this publication; therefore, we do not introduce any details.

Based on the test specification, one or more test bench configurations are derived. 
These test bench configurations are used to execute the specified test cases (which are part of the test procedures to be executed) using the considered test object.

\Figure[t!](topskip=0pt, botskip=0pt, midskip=0pt)[scale=0.95]{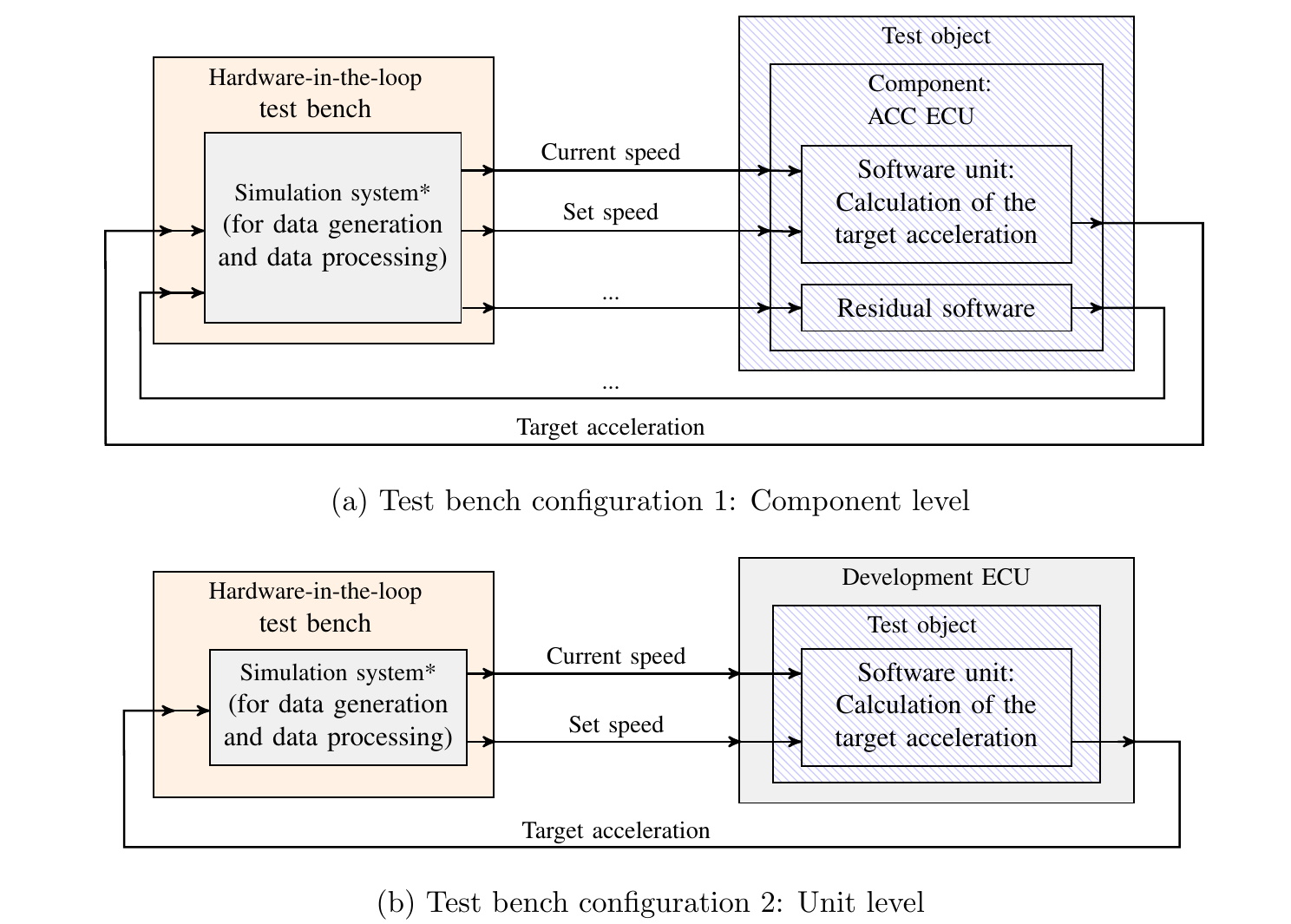}
{Example of two different test bench configurations (simplified depiction) at two different test levels (*configured to execute the test case; orange shaded: test bench; gray shaded: element provided at the test bench; blue hatched: test object; ECU: electronic control unit).\label{fig_TestBenchConfigurations}}

Fig.~\ref{fig_TestBenchConfigurations} (a) shows the \textbf{test bench configuration} derived to execute the test case ``Test Case \textit{SpeedControl}'' using the considered test object (simplified depiction).
The test object (hatched blue) is the component named the ``ACC ECU'' on which the (target) ACC software is running.  
This test bench configuration is provided at the \textbf{hardware-in-the-loop test bench} and includes a simulation system for data generation and data processing (which is configured to execute the test case; for example, selection and composition of necessary simulation models).
The simulation system's input and output interfaces must be connected to the hardware-in-the-loop test bench's interfaces; these test bench interfaces must be connected to the interfaces of the (target) ACC ECU on which the (target) ACC software is running. 
Thus, the (target) ACC ECU receives data from the hardware-in-the-loop test bench (in this example, the ego vehicle's current speed, the set speed, and other data necessary to operate the ACC ECU), processes these data, and returns the control values to the test bench (in this example, the target acceleration and other data generated by the ACC ECU).
The \textbf{elements} (shaded gray) required to execute the test case with this test bench configuration include, among others, the simulation system, which is configured to execute the test case.
The simulation system itself may consist of subelements such as different simulation models that are interconnected.
These subelements must be provided at the corresponding hardware-in-the-loop test bench (shaded orange).
Therefore, the elements that are part of this test bench configuration place requirements on the test bench used to execute the test cases.
Thus, the test bench configuration described above requires a hardware-in-the-loop test bench that is suitable for, among other things, running the simulation system for data generation and data processing as well as connecting to the ACC ECU.

To present a second example of a test bench configuration (based on the unit test plan, which is not described in detail in this publication), Fig.~\ref{fig_TestBenchConfigurations} (b) (simplified depiction) provides another example of a \textbf{test bench configuration} that can be used to execute the considered test case with a different test object.
The test object (hatched blue) is the software unit ``Calculation of the target acceleration'', which is running on a development ECU.
Compared to the test bench configuration described above, this test bench configuration requires an additional \textbf{element}: a development ECU capable of running this software unit.
Additionally, it must be possible to connect this development ECU to the test bench, which may impose additional requirements on the hardware-in-the-loop test bench used to execute this test case with this test object.

\subsubsection{Test Execution Phase}
\label{ssec_TestExecutionExample}

In this subsection, terms related to the test execution phase, which are introduced in Section~\ref{ssec_TestExecution}, are explained and distinguished from each other based on the development and testing of an ACC system. 
Fig.~\ref{fig_TestCaseExecution_UML} visualizes the proposed terms related to this test phase and their relationships as a UML diagram.

In this example, the first test bench configuration introduced in Section~\ref{ssec_TestConfigurationExample}, which is shown in Fig.~\ref{fig_TestBenchConfigurations} (a), is used to execute the test case ``Test Case \textit{SpeedControl}'' as specified in the test procedure specification.
During the execution of the test case, the test bench configuration~--~or, more precisely, the simulation system contained in the test bench configuration~--~generates and processes data during each simulation time step.
The simulation system runs on the hardware-in-the-loop test bench.
Based on the scene at time~$t_s$, the simulation system provides the current speed of the (simulated) ego vehicle, the set speed, and the other data related to operating the ACC ECU as \textbf{test object input values}.
These test object input values are provided to the ACC ECU through the test bench interfaces and thus stimulate the ACC system.
The ACC ECU processes these test object input values and returns the \textbf{test object output values} to the simulation system through the test bench interfaces.
In this example, the test object output values are the target acceleration and the other data generated by the ACC ECU.
The simulation system then processes these test object output values and generates the scene at time~$t_{s+1}$ based on the scene at time~$t_s$.
Thus, a closed-loop simulation is established.
This process is repeated in the next simulation time step until the concrete scenario ends. 
To evaluate the executed test case, the test bench configuration must generate the necessary data.
In this example, the \textbf{test case evaluation data} are recorded for a later test case evaluation and include, among other data, the data listed below: 

\begin{itemize}
	\item the simulation time, with a scene existing at each simulation time step $t_\text{s}$
	\item the state of the ACC system at each simulation time step~$t_\text{s}$
	\item the ego vehicle's speed at each simulation time step $t_\text{s}$
	\item the ego vehicle's (automatic) deceleration at each simulation time step $t_\text{s}$
\end{itemize}

Fig.~\ref{fig_TCED} illustrates the recorded test case evaluation data.

\Figure[t!](topskip=0pt, botskip=0pt, midskip=0pt)[]{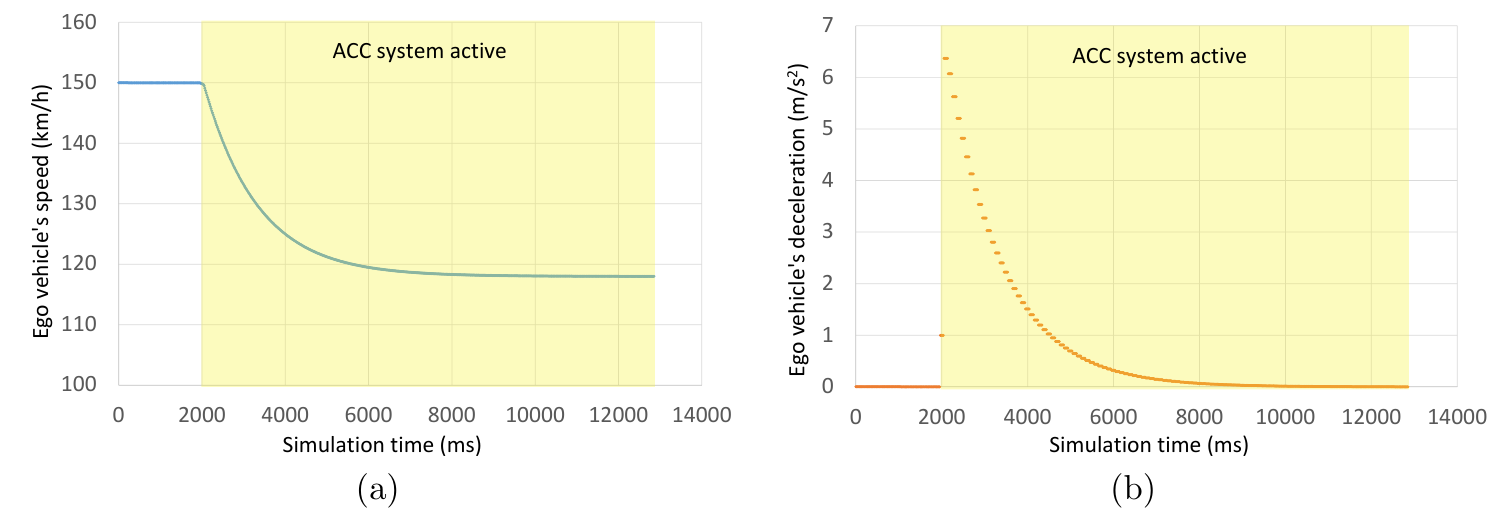}
{Test case evaluation data generated during the execution of the test case (data were generated using a prototypically implemented ACC system connected to a simulation with nonvalidated simulation models to illustrate both the example and the proposed terms).\label{fig_TCED}}

\subsubsection{Test Evaluation and Reporting Phase}

In this subsection, terms related to the test evaluation and reporting phase, which are introduced in Section~\ref{ssec_TestEvaluation}, are explained and distinguished from each other based on the development and testing of an ACC system. 
Fig.~\ref{fig_TestEvaluation_UML} visualizes the proposed terms related to this test phase and their relationships as a UML diagram.

In this example, the \textbf{test evaluation tool} evaluates the two evaluation criteria belonging to the executed test case introduced in Section~\ref{ssec_TestSpecificationExample}.
For this purpose, metric results are calculated based on test case evaluation data according to the two evaluation metrics as soon as the application periods are active.
Fig.~\ref{fig_EV} illustrates the calculated metric results.

\Figure[t!](topskip=0pt, botskip=0pt, midskip=0pt)[]{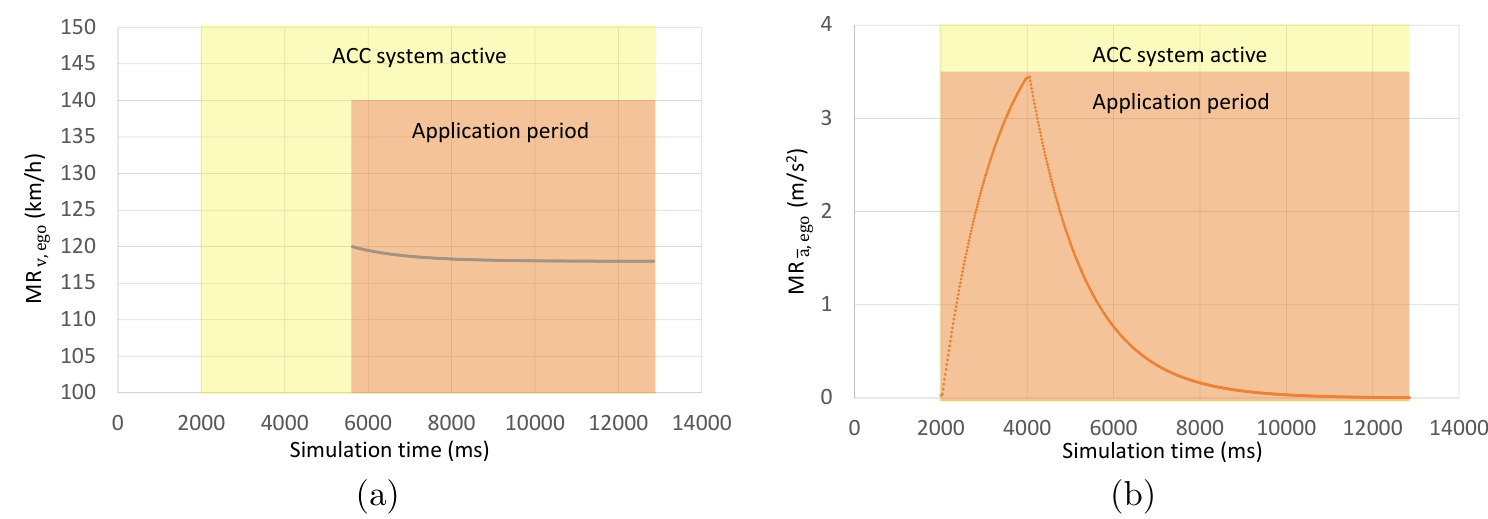}
{Metric results calculated based on the two evaluation metrics named (a) ``Ego\_speed'' ($\text{MR}_\text{v,ego}$) and (b) ``Average\_ego\_deceleration'' ($\text{MR}_\text{${\bar a}$,ego}$) within the application period.\label{fig_EV}}

When applying the first evaluation criterion, the test evaluation tool evaluates whether all of the calculated metric results are maintained within a maximum deviation of $\SI{-5}{\percent}$ from the set speed.
Additionally, the corresponding evaluation scale is applied.
As shown in Fig.~\ref{fig_EV}~(a), the evaluation criterion is fulfilled for all metric results.

When applying  the second evaluation criterion, the test evaluation tool evaluates whether all of the calculated metric results are less than or equal to \SI{3,5}{\meter \per \second\squared}.
As shown in Fig.~\ref{fig_EV}~(b), the evaluation criterion is fulfilled for all metric results.

Based on all individual evaluations, a \textbf{test report} is generated.
If all evaluation criteria are fulfilled, the executed test case is considered to be passed, which is the case in this example. 
Using the evaluation scale, it is also possible to make a statement about how well this test case is passed.

\section{Evaluation of the requirements set for the taxonomy} \label{sec_req_evaluation}

We created the current state of the taxonomy~--~consisting of the terms that are part of the proposed basic vocabulary, their descriptions, and the relationships between these terms~--~based on the requirements listed in Section~\ref{sec_requirements}.
In this section, we evaluate the fulfillment of these requirements.
Since this publication does not aim to develop a complete taxonomy, we evaluate only the set requirements concerning the current state of the taxonomy.
As already mentioned, the current state of the taxonomy will be extended in the future. 
In doing so, the requirements mentioned above will continue to be considered.

We created the current state of the taxonomy based on a review of relevant literature. 
For this purpose, we analyzed various publications.
There are many different standards (Req.~1\,a), most of which are complementary and/or deal with specific topics within development and test processes.
Therefore, we focused on the ISO/IEC/IEEE~29119~standard~\cite{ISO29119_2013} since it bundles and replaces different standards. 
Furthermore, the ISO/IEC/IEEE~29119~standard~\cite{ISO29119_2013} provides an overview of the entire test process.
Regarding the prevailing parlance in the scientific literature (Req.~1\,b), we focused mainly on ISTQB activities since the ISTQB has contributed to harmonizing the progressively developed terms in the extensive literature regarding testing.
Regarding the colloquial language of both researchers and developers (Req.~1\,c), we also focused on the ISTQB activities since those who have obtained an ISTQB certificate as software testers are familiar with the terms in the ISTQB syllabus and their definitions. 
However, the extent to which these and other people use the terms in their colloquial language is something we can judge only to a limited extent.
Our experience with scenario-based development and test approaches as well as conversations with various people have allowed us to consider the colloquial language of those we spoke with.
In summary, we aimed to be as compliant as possible with existing standards, with the prevailing parlance in the scientific literature, and with the colloquial language of researchers and developers.
However, the descriptions of the proposed terms are based primarily on existing standards, as we consider standards to be particularly relevant.
For the current state of the taxonomy, we consider Req.~1\,a, Req.~1\,b, and Req.~1\,c to be fulfilled to the degree to which we can evaluate them without extensive feedback from the community.

We clearly structured the taxonomy by dividing it into different phases within a development and test process (Req.~2\,a).
Due to input from the people we conversed with and the feedback we received due to the prepublication of an earlier version of this publication on arXiv, the current state of the taxonomy is clearly formulated (Req.~2\,b) and defined as simply as possible (Req.~2\,c).
However, we cannot conclusively evaluate whether these requirements are sufficiently fulfilled for other people without extensive feedback from the community.
Therefore, we consider Req.~2\,a, Req.~2\,b, and Req.~2\,c to be fulfilled to the degree to which we can evaluate them without extensive feedback from the community.

By representing the relationships between the proposed terms as UML diagrams, we were able to resolve and avoid inconsistencies and contradictions within the taxonomy, which is why we consider Req.~2\,d to be fulfilled.
This form of representing the taxonomy also allowed us to visualize the terms that are part of the taxonomy and their relationships (i.e., the current state of the taxonomy) in a clear, concise, and understandable way.
For this reason, we also consider Req.~3 to be fulfilled.

A researcher or developer can adapt terms and relate and contextualize additional terms that are relevant to him or her to existing terms and thus add these terms to the taxonomy.
Furthermore, he or she can adapt the relationships between terms visualized as UML diagrams.
For example, Hoßbach~\cite{Hossbach_2019} has already extended the taxonomy we proposed in~\cite{Steimle_2018}, on which this publication is based, to include terms relevant to him.
Therefore, the taxonomy is extensible and adaptable, which is why we also consider Req.~4 to be fulfilled.

Since the current state of the taxonomy includes only those terms and their relationships that we consider particularly relevant to an overview of these approaches, the current state of the taxonomy is intended to be as universally applicable as possible.
For this reason, we also consider Req.~5, which relates especially to this publication, to be fulfilled.

As described above, we cannot conclusively evaluate the fulfillment of all requirements identified without feedback from the community. 
Therefore, we welcome any feedback regarding the current state of the taxonomy.
Due to the prepublication of an earlier version of this publication on arXiv, we were already able to discuss and consider the feedback we received when creating this publication.

\section{Conclusion and Outlook} \label{sec_conclusion}

This publication contributes to a consistent taxonomy for scenario-based development and test approaches for automated vehicles.
Based on the taxonomic requirements we have identified, we proposed a framework to maintain clarity and provide structure.
The framework is based on the phases of a V-model development and test process.
Thus, we proposed structuring the taxonomy into several phases: scenario derivation and usage; the concept, design, and implementation phase; and the test process. 
As this publication’s focus is on scenario-based test approaches, we detailed the test process according to the test phases contained therein: the test planning phase; the test design and implementation phase; the test configuration phase; the test execution phase; and the test evaluation and reporting phase.

Based on the proposed framework and the taxonomic requirements we have identified, we proposed and described terms that we consider particularly relevant to an overview of scenario-based development and test approaches for automated vehicles.
These terms represent a basic vocabulary that will be extended in the future. 
The terms of each phase and their relationships were visualized as UML diagrams. 
This visualization enables the quick recognition of relationships and dependencies between different terms.

We applied the proposed basic vocabulary by explaining and distinguishing the proposed terms based on the development and testing of an ACC system. 
For this purpose, we provided examples for each proposed term and described the relationships between the terms.

Finally, we evaluated the fulfillment of the requirements set for the taxonomy.

In the future, the current state of the taxonomy will be discussed and extended. 
Visualization using UML diagrams allows researchers and developers in the automotive domain to relate and contextualize additional terms and thus add these terms to the taxonomy.
Additionally, this visualization allows him or her to identify relationships, inconsistencies, inaccuracies, and contradictions within the extended taxonomy.
Whether visualization of the taxonomy using UML diagrams is still possible if many more terms are included in the taxonomy and whether this form of visualization and the proposed framework are still useful will be evaluated in the future.

Regarding the proposed framework, basic vocabulary, presentation style, and extension of the taxonomy, we welcome any feedback from the community.
Additionally, we hope that researchers and developers will extend the basic vocabulary we proposed in the future and that the basic vocabulary and presentation style, in particular, will help them in their daily work.%

\section*{Acknowledgments}

We want to thank Günter Ehmen (OFFIS e.V.) for feedback on the presentation of the UML diagrams.
We also thank Julian Fuchs, Felicitas Kunz, Christian Steinhauser (FZI Research Center for Information Technology), and Nils Müllner (DLR~e.V.) for detailed feedback on the prepublication on arXiv.

\bibliographystyle{IEEEtran}%
\bibliography{literature/literature}%

\begin{IEEEbiography}[{\includegraphics[width=1in,height=1.25in,clip,keepaspectratio]{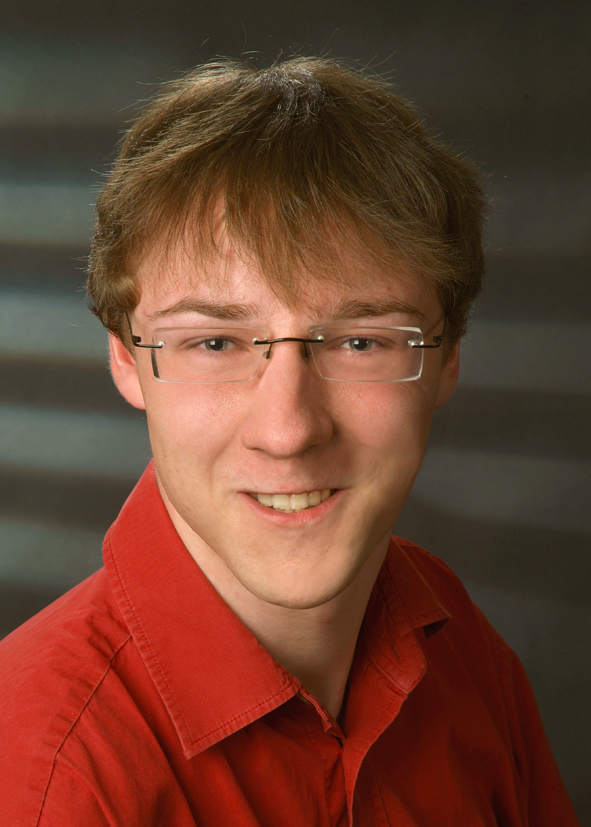}}]{Markus Steimle} received a B.Eng. degree in electrical engineering and information technology from the Landshut University of Applied Sciences, Landshut, Germany, in 2014. 
In 2016, he received an M.Sc. degree in automotive software engineering from the Technical University of Munich, Munich, Germany.
Since 2016, he has been a Ph.D. student at the Institute of Control Engineering of Technische Universität Braunschweig, Braunschweig, Germany. 
His main research interests are scenario-based verification and validation of automated vehicles, focusing on the use of simulative testing methods.
\end{IEEEbiography}

\begin{IEEEbiography}[{\includegraphics[width=1in,height=1.25in,clip,keepaspectratio]{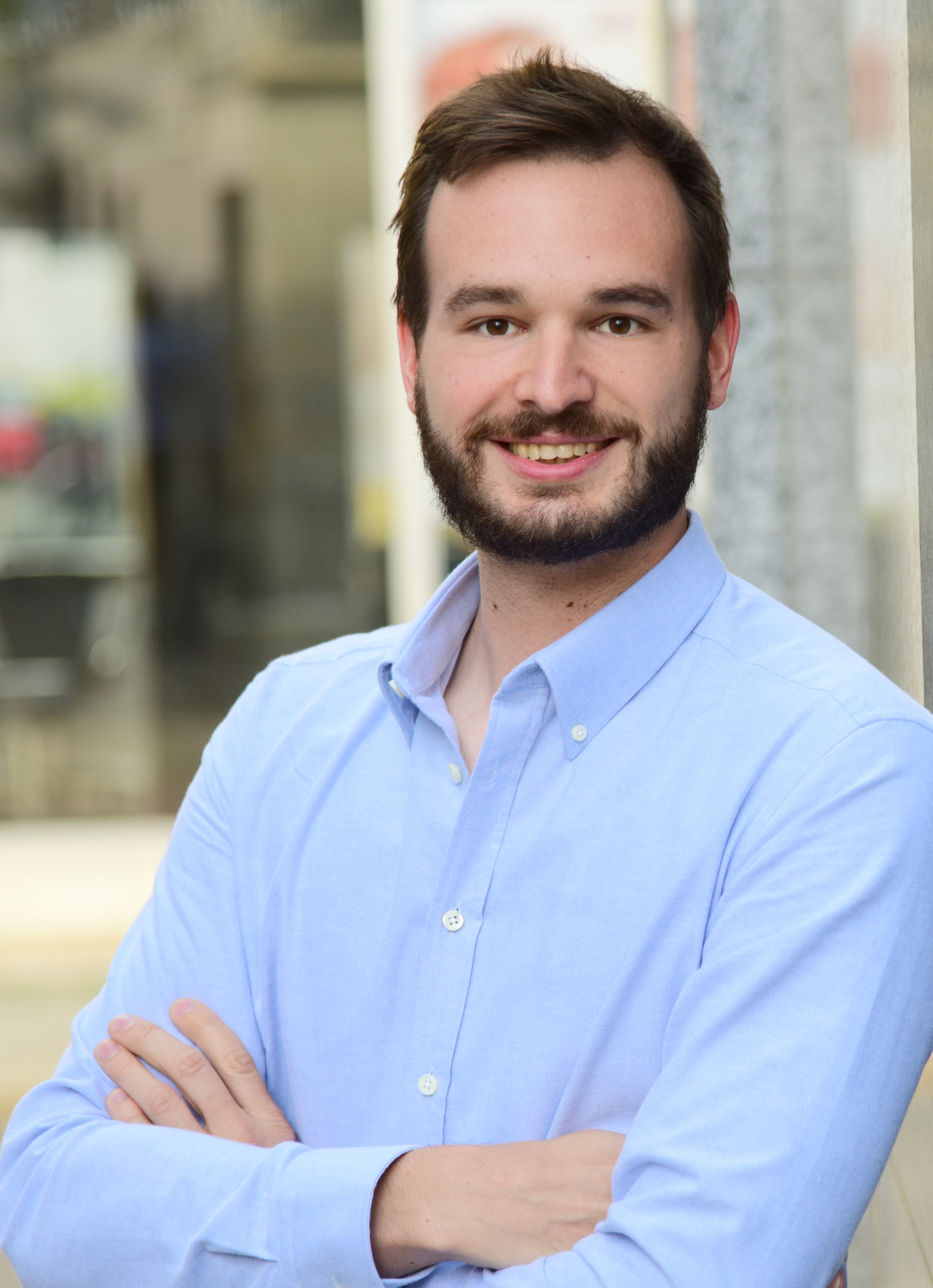}}]{Till Menzel} received a B.Sc. and M.Sc. degrees in electrical engineering from the Technische Universität Braunschweig, Braunschweig, Germany, in 2011 and 2014 respectively.
Since 2014, he has been a Ph.D. student at the Institute of Control Engineering of Technische Universität Braunschweig, Braunschweig, Germany. 
His main research interests are scenario-based verification and validation of automated vehicles, focusing on a systematic generation of scenarios for simulation-based testing.
\end{IEEEbiography}

\begin{IEEEbiography}[{\includegraphics[width=1in,height=1.25in,clip,keepaspectratio]{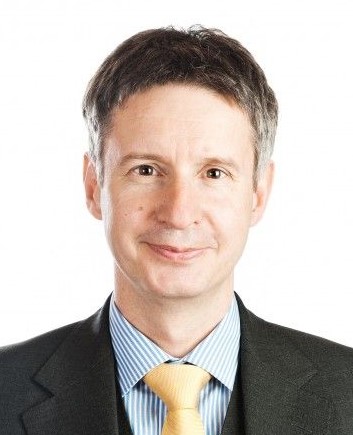}}]{Markus Maurer} studied Electrical Engineering at the Technische Universität München (Diplom 1993).
He then joined the group of Prof. E.\,D. Dickmanns at the Universität der Bundeswehr München where he finished his Ph.D. in 2000 in the field of automated driving. 
From 1999 to 2007 Prof. Maurer was a project manager and head of the development department of Driver Assistance Systems at Audi. 
Since 2007 he has been a full professor for Automotive Electronics Systems at the Institute of Control Engineering at Technische Universität Braunschweig.
His research focuses on both functional and systemic aspects of automated road vehicles. 
(Picture: Jessen Oestergaard / Daimler und Benz Stiftung)
\end{IEEEbiography}

\EOD

\end{document}